\documentclass[prd,showpacs,superscriptaddress,twocolumn,floatfix,nofootinbib,preprintnumbers,altaffilletter, letterpaper]{revtex4-1}

\usepackage{amsmath,amssymb}
\usepackage{xspace}
\usepackage{xcolor}
\usepackage{graphicx}
\usepackage[caption=false]{subfig}
\usepackage{multirow}
\usepackage{enumitem}
\graphicspath{{.}{./images/}}
\usepackage{color}
\usepackage[acronym,shortcuts]{glossaries}
\usepackage{times}
\usepackage{bm}
\usepackage{array}
\usepackage{colortbl}
\usepackage{booktabs}
\usepackage[colorlinks=true,citecolor=teal]{hyperref}
\usepackage[T1]{fontenc}

\newcommand{\orcid}[1]{\href{https://orcid.org/#1}{\includegraphics[width=10pt]{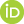}}}

\newcommand{\mlstat}{\mbox{MLStat}\xspace}
\newcommand{\pycbc}{{\sc PyCBC}\xspace}
\newcommand{\Pcbc}{{$\mathrm{P}_\mathrm{CBC}$}\xspace}
\newcommand{\msun}{\text{M}_{\odot}\xspace}
\newcommand{\inj}{\texttt{INJ}\xspace}
\newcommand{\cleaninj}{\texttt{CLEANINJ}\xspace}
\newcommand{\gn}{\texttt{GN}\xspace}
\newcommand{\gl}{\texttt{GL}\xspace}
\newcommand{\D}{\mbox{{$\mathcal{D}_\theta$}}}
\newcommand{\G}{\mbox{{$\mathcal{G}_\phi$}}}
\newcommand{\overbar}[1]{\mkern 1.5mu\overline{\mkern-1.5mu#1\mkern-1.5mu}\mkern 1.5mu}
\newcommand{\injtest}{\texttt{INJ-TEST}\xspace}
\newcommand{\gltest}{\texttt{GL-TEST}\xspace}
\newcommand{\trigtest}{\texttt{TRIG-TEST}\xspace}

\newacronym{mle}{MLE}{maximum likelihood estimation}
\newacronym{ml}{ML}{machine learning}
\newacronym{roc}{ROC}{Receiver operating characteristic}
\newacronym{auc}{AUC}{area under the curve}
\newacronym{gw}{GW}{gravitational-wave}
\newacronym{far}{FAR}{false-alarm rate}
\newacronym{kl}{KL}{Kullback-Leibler}
\newacronym{tsne}{t-SNE}{t-distributed Stochastic Neighbour Embedding}
\newacronym{pn}{PN}{post-Newtonian}
\newacronym{psd}{PSD}{power-spectral density}
\newacronym{snr}{SNR}{signal-to-noise ratio}
\newacronym{dl}{DL}{deep learning}
\newacronym{cwt}{CWT}{continuous wavelet transforms}
\newacronym{cbc}{CBC}{compact binary coalescence}
\newacronym{bbh}{BBH}{binary black-hole}
\newacronym{imbh}{IMBH}{intermediate-mass black hole}
\newacronym{vae}{VAE}{variational auto-encoder}
\newacronym{gan}{GAN}{generative adversarial network}
\newacronym{pca}{PCA}{Principal Component Analysis}
\newacronym[plural=GPUs]{gpu}{GPU}{graphics processing unit}
\newcommand{\mle}{\ac*{mle}\xspace}
\newcommand{\ml}{\ac*{ml}\xspace}
\newcommand{\roc}{\ac*{roc}\xspace}
\newcommand{\auc}{\ac*{auc}\xspace}
\newcommand{\gw}{\ac*{gw}\xspace}
\newcommand{\far}{\ac*{far}\xspace}
\newcommand{\kl}{\ac*{kl}\xspace}
\newcommand{\tsne}{\ac*{tsne}\xspace}

\newcommand{\psd}{\ac*{psd}\xspace}
\newcommand{\dl}{\ac*{dl}\xspace}
\newcommand{\snr}{\ac*{snr}\xspace}
\newcommand{\cwt}{\ac*{cwt}\xspace}
\newcommand{\cbc}{\ac*{cbc}\xspace}
\newcommand{\bbh}{\ac*{bbh}\xspace}
\newcommand{\imbh}{\ac*{imbh}\xspace}
\newcommand{\vae}{\ac*{vae}\xspace}
\newcommand{\gan}{\ac*{gan}\xspace}
\newcommand{\pca}{\ac*{pca}\xspace}
\newcommand{\gpu}{\ac*{gpu}\xspace}

\begin{document}

\title{Towards a robust and reliable deep learning approach for detection of compact binary mergers in gravitational wave data}

\author{Shreejit Jadhav\orcid{0000-0003-0554-0084}}
\email{spjadhav@swin.edu.au}
\affiliation{Inter-University Centre for Astronomy and Astrophysics (IUCAA), Post Bag 4, Ganeshkhind, Pune 411 007, India}
\affiliation{Centre for Astrophysics and Supercomputing, Swinburne University of Technology, Hawthorn, Victoria 3122, Australia}
\affiliation{ARC Centre of Excellence for Gravitational Wave Discovery (OzGrav), Melbourne, Australia}
\author{Mihir Shrivastava\orcid{0000-0002-6767-139X}} 
\email{mihirshriv@iitkgp.ac.in}
\affiliation{Indian Institute of Technology (IIT), Kharagpur, India}
\author{Sanjit Mitra\orcid{0000-0002-0800-4626}}
\email{sanjit@iucaa.in}
\affiliation{Inter-University Centre for Astronomy and Astrophysics (IUCAA), Post Bag 4, Ganeshkhind, Pune 411 007, India}
\date{\today}

\begin{abstract}
The ability of deep learning (DL) approaches to learn generalised signal and noise models, coupled with their fast inference on GPUs, holds great promise for enhancing gravitational-wave (GW) searches in terms of speed, parameter space coverage, and search sensitivity. However, the opaque nature of DL models severely harms their reliability. In this work, we meticulously develop a DL model stage-wise and work towards improving its robustness and reliability. First, we address the problems in maintaining the purity of training data by deriving a new metric that better reflects the visual strength of the ``chirp'' signal features in the data. Using a reduced, smooth representation obtained through a variational auto-encoder (VAE), we build a classifier to search for compact binary coalescence (CBC) signals. Our tests on real LIGO data show an impressive performance of the model. However, upon probing the robustness of the model through adversarial attacks, its simple failure modes were identified, underlining how such models can still be highly fragile. As a first step towards bringing robustness, we retrain the model in a novel framework involving a generative adversarial network (GAN). Over the course of training, the model learns to eliminate the primary modes of failure identified by the adversaries. Although absolute robustness is practically impossible to achieve, we demonstrate some fundamental improvements earned through such training, like sparseness and reduced degeneracy in the extracted features at different layers inside the model. We show that these gains are achieved at practically zero loss in terms of model performance on real LIGO data before and after GAN training. Through a direct search on $\sim 8.8$ days of LIGO data, we recover two significant CBC events from GWTC-2.1~\cite{gwtc2.1}, GW190519\_153544 and GW190521\_074359. We also report the search sensitivity obtained from an injection study.

\end{abstract}

\maketitle

\section{Introduction}\label{ch:RDL:sec:intro}

    The first detection of a \gw signal originating from a \bbh merger~\cite{Abbott2016b} by the advanced LIGO detectors~\cite{Aasi2015} marked the beginning of a new era in astronomy. Collectively, more than ninety \gw events have been detected so far~\cite{Nitz2018a,Nitz2020,Nitz2021,Abbott2019,Abbott2021,gwtc2.1,gwtc3,Venumadhav2019} in the data released~\cite{Rich_Abbott_2021,Vallisneri_2015} by the advanced LIGO~\cite{Aasi2015} and advanced Virgo~\cite{Acernese2015} detectors. These detections include a variety of \cbc events of great astrophysical implications~\cite{Abbott2016,Abbott2016b,Abbott2016c,Abbott2017a,Abbott2017c,Abbott2017b,Abbott2017d,Abbott2020,Abbott2020a,Abbott2020b,Abbott2020c,Zackay2019,Abbott2016a,Abbott_2021}. Advanced LIGO detectors and KAGRA~\cite{Aso2013,Akutsu2020} began the fourth observing run on May $24$, $2023$ and Virgo is planning to join shortly~\cite{O4runplan}. In the latter half of the decade, a third LIGO detector in India is expected to join the network~\cite{IndIGO_LIGO_INDIA,LIGO_india,Abbott2018}, which will enable improvements in the event significance, polarisation studies, sky localisation and overall duty factor~\cite{Saleem_2021}.

    In the past few years, applications of machine learning algorithms have been explored for a variety of tasks in \gw data analysis, most notably, \cbc searches~\cite{Schafer_one2many,Schafer2020_BNS,Men_ndez_V_zquez_2021,2022arXiv221101520N}, sensitivity improvements of existing search pipelines~\cite{Jadhav_2021,2022PhRvD.105h3018M,2022PhRvD.105j4056M,2021PhRvD.104b3014M,Kapadia_2017,2015CQGra..32x5002K}, non-linear noise modelling and subtraction~\cite{2021arXiv211103295Y,2020PhRvR...2c3066O} that is useful for early alerts for \cbc{s}~\cite{2021PhRvD.104f2004Y}, parameter estimation~\cite{2022NatPh..18..112G,Green2020}, glitch classification~\cite{Mukund2017,2020PhRvD.102h4034B,2015CQGra..32u5012P,2017CQGra..34c4002P,Glanzer2022,Choudhary2022,Soni2021}, construction of population models~\cite{2019MNRAS.488.3810P,PhysRevD.98.083017,PhysRevD.101.123005,PhysRevD.106.103013,ruhe2022normalizing,Riley_2023}, approximate \gw detectability~\cite{Gerosa_2020,wong2020gravitationalwave,Talbot_2022,ChapmanBird_2023} and searches for other interesting signals originating from \imbh mergers~\cite{thames_22}, strongly-lensed events~\cite{2021PhRvD.104l4057G}, short gamma-ray bursts~\cite{2015CQGra..32x5002K}, continuous waves~\cite{2020PhRvD.102h3024B} and unknown astrophysical events~\cite{Moreno2022,PhysRevD.105.083007}.

    This work focuses on the applications of \dl for \cbc searches. Though template-based matched-filtering searches are presently the most sensitive, their parameter space coverage is limited by the availability of computational resources~\cite{PhysRevD.60.022002}. This limitation also restricts the search sensitivity for astrophysically interesting parameters like eccentricity, precession and higher-order modes. While the unmodelled searches do not have such restrictions on the expected signal, they are relatively less sensitive due to a loosely bounded likelihood function and, more importantly, the presence of glitches in the data. The strength of \dl models in distinguishing glitches from \cbc signals makes them excellent alternative tools for searches~\cite{Jadhav_2021,Schafer_one2many,Schafer2020_BNS,Men_ndez_V_zquez_2021,thames_22}. Also, \dl methods readily utilise accelerated computing hardware like \glspl{gpu}, improving the speed by orders of magnitude.

    The field of \gw searches has witnessed promising developments on the \ml front in recent years. Some of the early works professing the advantages of \ml methods for \gw searches focused on sensitivity comparisons against matched-filtering primarily using simulated noise~\cite{George2018,Gabbard2018} with a limited extension to real data~\cite{George2018a}. These works based their sensitivity estimates on \roc curves obtained from datasets consisting of discrete samples that placed \cbc injections in specific regions of the analysis window. \citet{Gebhard2019} critically examined the suitability of such approaches for realistic searches by addressing various issues like class imbalance, temporal localisation of candidate events and assigning statistical significance. A more reliable approach of performing trigger-based coincident analysis with the \dl outputs on a stream of continuous data was proposed in the literature~\cite{Schafer2020_BNS,Schafer_one2many}, which is followed in this work as well. A systematic approach in constructing a proper ranking statistic is also a crucial step towards implementing a coincident search~\cite{Jadhav_2021,Kim2020,2021PhRvD.104b3014M,Schafer_one2many}. To enable comparative evaluation of different \ml pipelines in a realistic search setting and to provide a common platform, a mock data challenge was organised recently~\cite{Schafer2022_mlgwsc}.

    The \dl approaches come with a cost. The most critical hurdle that keeps them from being considered for production-level deployment is the unreliability originating from their black-box nature. It has been shown that tiny perturbations to the input images achieved through targeted attacks can lead to surprise failures of these models~\cite{goodfellow2014explaining}. Also, generative adversarial attacks can reveal many other failure modes. A generator trained to deceive a \dl model can yield many unrealistic images which do not belong to any meaningful class, although, the model classifies them into specific classes with high probabilities. These failures highlight the ambiguity over the actual features learnt by a \dl model that otherwise shows an excellent performance in metrics like accuracy, F1-score, and area under the \roc curve. It also highlights how the performance of \dl models on new data is unpredictable and fragile, making them unreliable for the analysis of real data at the production stage. In the context of \gw searches, by obtaining adversarial examples through input perturbations, \citet{Gebhard2019} first demonstrated the simple ``failure modes'' of \dl models, which the conventional approaches could handle easily. In this work, we take a \gan-based approach to generate the adversarial samples. \gan{s} have previously been used in \gw data analysis primarily for generating various transient signals~\cite{2022arXiv220700207P,2021CQGra..38o5005M}. Here, we take a novel approach of utilising the \gan framework for bringing robustness to our \dl model.

    In this work, we explore several ways of making \dl models more reliable. Although the focus is on \cbc searches, the methods presented here are extendable to searches for signals from other sources and possibly to tasks other than searches, such as glitch classification and parameter estimation. In Section~\ref{ch:RDL:sec:showing}, we address the issues related to the purity of the training data, describe the development of a new metric for data generation and the preparation of various datasets for different stages of training. Section~\ref{ch:RDL:sec:looking} elaborates the training of a \vae model to achieve a reduced representation of the signal-specific features in the input. Repurposing this information for the task of signal identification is detailed in Section~\ref{ch:RDL:sec:utility}. Section~\ref{ch:RDL:sec:robustness} describes the scrutiny of the resulting classifier model for robustness and its retraining in the \gan setup to increase robustness. Lastly, the implementation of a direct search using the classifier is provided in Section~\ref{ch:RDL:sec:search}. We conclude with outlook and future directions in Section~\ref{ch:RDL:sec:concl}.

\section{What are we showing to the model?}\label{ch:RDL:sec:showing}

    In this section, we describe the development of a new metric for data generation and the preparation of datasets. For efficient training of various stages of our model, we construct four different datasets for training: injections in simulated Gaussian noise (\inj), the same set of injections without noise (\cleaninj), the same Gaussian noise realisations without injections (\gn) and a glitch dataset (\gl) obtained from GravitySpy~\cite{Zevin2017,Bahaadini2018,Glanzer2022,Soni2021}. To ensure near absolute fidelity of the training data, we stick to simulated Gaussian noise. However, we consider only real LIGO data for testing. We prepare three classes of test datasets: \cbc injections (\injtest), \pycbc background triggers (\trigtest) and glitches (\gltest). Our experiments on the test datasets show that the features learnt by our model are generalised enough to ensure the model's applicability to real data.

\subsection{New metric for injection data}

    \begin{figure}[htb]
        \subfloat[$\mathcal{M} = 9.0$; $\rho = 8.2$]
        {%
            \centering
            \includegraphics[width=0.48\linewidth, trim=0.8cm 0cm 2cm 2cm]{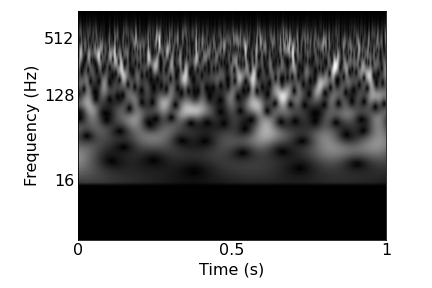}
        }\hfill
        \subfloat[$\mathcal{M} = 54.3$; $\rho = 8.1$]
        {%
            \centering
            \includegraphics[width=0.48\linewidth, trim=1.8cm 0cm 1cm 2cm]{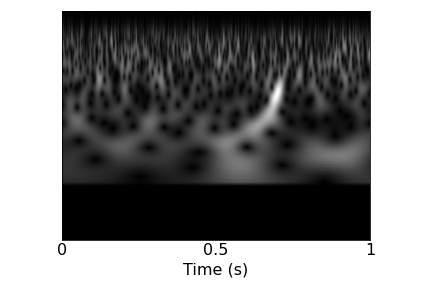}
        }
        \caption{Example CWT maps of two injections highlighting how the chirp-masses of \cbc signals determine their \emph{visibility} in the time-frequency domain representation even when the recovered SNR values are nearly the same. However, in $\gamma$ metric (as described in Equation~\ref{ch:RDL:eq:gamma}), these injections are well separated with values $26.5$ and $44.9$, respectively.}
        \label{ch:RDL:fig:visibility_difference}
    \end{figure}

    Most of the past works considered a fixed lower cutoff on \snr across the parameter space for generating training data either directly or by fixing the upper cutoff on the chirp-distance\footnote{Chirp-distance is the luminosity distance adjusted for the leading order chirp mass term in the \cbc waveform.}. However, the duration of time the signal remains in the sensitive band of a detector, which in turn depends on the component masses and the \psd, strongly determines the \emph{visibility} of the signal\footnote{Here, \emph{visibility} refers to the relative strength of the pixels corresponding to the signal features in the input image to those corresponding to the noise features. It can also be imagined as a metric that captures humans' ability to distinguish signal features from noise features in the image data.}. Figure~\ref{ch:RDL:fig:visibility_difference} illustrates this with the \cwt maps of two injections that have nearly the same recovered \snr values but different chirp masses. Clearly, the low chirp-mass binary has a smaller amplitude but spends a much longer time in the sensitive band of LIGO/Virgo. Therefore, the total energy contained in its signal is distributed over a much longer duration resulting in reduced visibility of its chirp features compared to the high chirp-mass binary. This difference is expected to play a crucial role for data representations which do not include the phase information, e.g., time-frequency representations such as \cwt maps, Q-transforms, and short-time Fourier transforms saved as images with pixel values proportional to the amplitude. 
    
    To efficiently train the \dl models to detect the weakest possible signals across the parameter space, maintaining uniform visibility of signal features across the training data is important. Therefore, instead of using \snr-based thresholds, an approximate metric that better reflects the signal's visible strength could be the \snr divided by the duration of the signal in the detector's sensitive band. For a detector with flat frequency response, the duration of the \cbc signal can be captured in leading order using the \emph{chirp time} $(\tau_0)$. It is the Newtonian coalescence time calculated starting from the frequency $f_0$ and is given by,
    \begin{equation}
        \tau_0 = \frac{5}{256 \pi f_0} \bigg( \frac{\pi G \mathcal{M} f_0}{c^3} \bigg)^{-5/3} \,.
    \end{equation}
    Considering that the detector noise is coloured, we can choose $f_0$ in such a way that we operate in a frequency band where the response of the detector is approximately flat and $\tau_0$ remains relevant even after whitening. However, we found that the approximate metric constructed this way lost its validity rapidly upon approaching higher masses where Newtonian approximation fails to describe the waveform. To solve this issue, we numerically modelled the intrinsic mass dependence of the visibility of the chirp features. We obtained maximum amplitude values of the normalised whitened waveforms generated at each point in the component mass space to capture the variations in the maximum pixel values corresponding to the \cbc chirp features in the \cwt maps. We uniformly divided the mass-ratio ($q$) - total mass ($M$) space\footnote{All the masses mentioned here are in the detector frame.}, generated waveforms at each point of the grid using \texttt{IMRPhenomXPHM} approximant and normalised them (such that $\rho(\hat{\mathrm{\textbf{h}}}, \hat{\mathrm{\textbf{h}}}) = 1$, where $\hat{\mathrm{\textbf{h}}}$ is the normalised waveform). The $q$ and $M$ values were sampled in the range $[1,8]$ and $[4,800]$ with step-sizes $0.1$ and $10$, respectively. We whitened both '$+$' and '$\times$' polarisations of each waveform using the \texttt{aLIGOZeroDetHighPower} \psd, and recorded the maximum amplitude of their quadrature sum as the intrinsic mass dependence factor in the amplitude, $A_\mathrm{intr}(q_i, M_i)$ for the $i^\mathrm{th}$ binary. 

    \begin{figure}
        \centering
        \includegraphics[clip=true, trim=1cm 0cm 2cm 1cm, width=\linewidth]{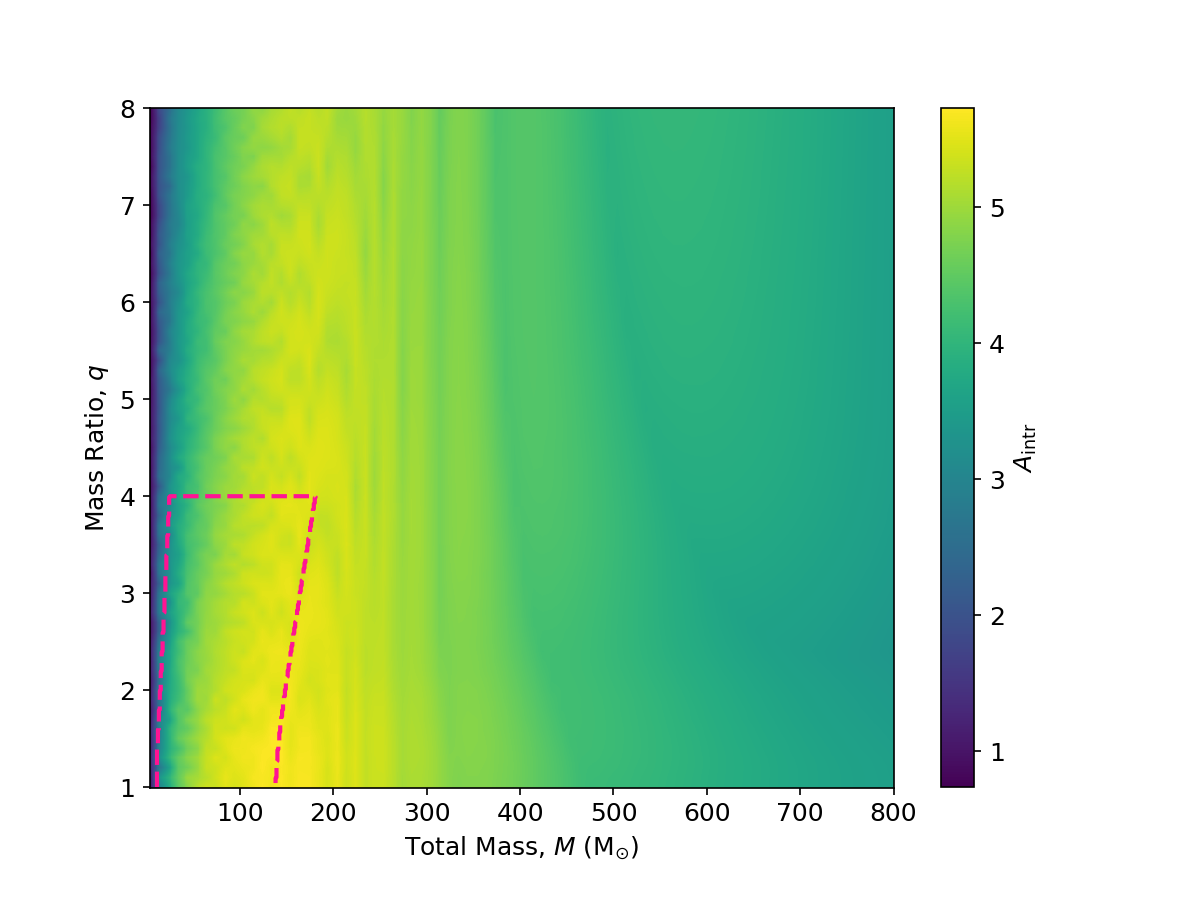}
        \caption{Value of the maximum amplitude of normalised whitened waveform $A_\mathrm{intr}(q, M)$ as a function of the mass ratio $q$ and total mass $M$. We use \texttt{aLIGOZeroDetHighPower} PSD for whitening. The relative variation of $A_\mathrm{intr}(q, M)$ can be as large as $\sim 7$ in the extreme regions of the mass space. The dashed line marks the region of parameter space used for the training and testing.}
        \label{ch:RDL:fig:A_intr_contour}
    \end{figure}

    We used linear interpolation between the grid points to obtain a continuous function $A_\mathrm{intr}(q, M)$ that is valid at all points within the range of parameters considered for this numerical estimation. It can be observed from Figure~\ref{ch:RDL:fig:A_intr_contour} that the maximum relative variation in $A_\mathrm{intr}(q, M)$ is as high as $\sim 7$. Using this intrinsic mass dependence, we construct a new metric that better reflects the \emph{visible strength} of the chirp features. We call this metric $\gamma$, which is obtained as,
    \begin{equation} \label{ch:RDL:eq:gamma}
        \gamma(\boldsymbol\theta) = A_\mathrm{intr}(\boldsymbol\theta)  \rho(\mathrm{\textbf{s}}, \hat{\mathrm{\textbf{h}}}(\boldsymbol\theta)),
    \end{equation}
    where, $\rho(\mathrm{\textbf{s}}, \hat{\mathrm{\textbf{h}}}(\boldsymbol\theta))$ is the \snr calculated with data \textbf{s} and template $\hat{\mathrm{\textbf{h}}}(\boldsymbol\theta)$ with $\boldsymbol\theta$ limited to only the mass parameters.

    \begin{figure*}
        \centering
        \includegraphics[clip=true, trim=6cm 9cm 6cm 8cm, width=0.9\textwidth]{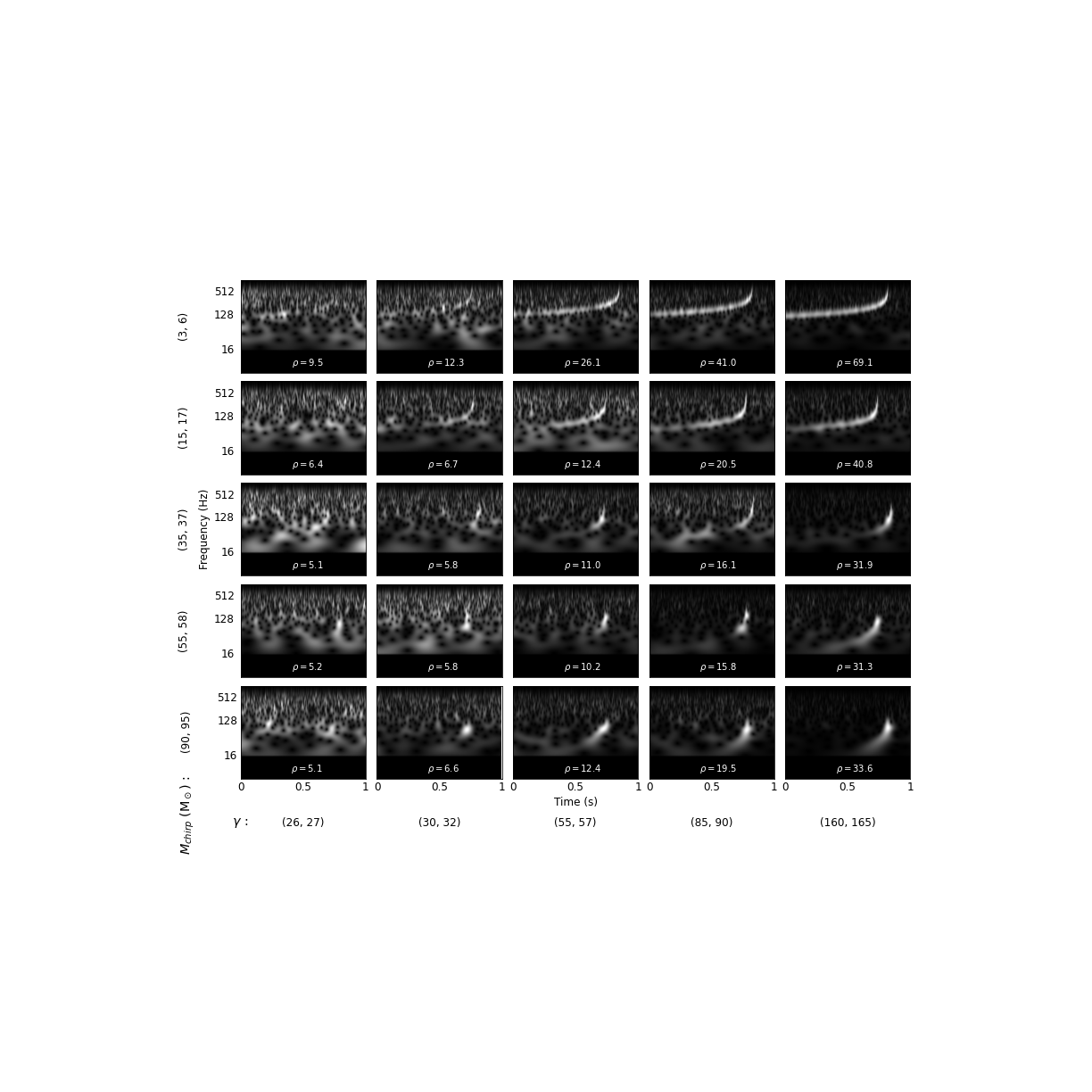}
        \caption{Variation in the visible strength of chirp features with the metric $\gamma(\theta)$, independent of the chirp mass. The images are sampled from the \inj dataset with $M_{chirp}$ and $\gamma$ values lying in the corresponding bins shown in the brackets. Recovered \snr values are annotated on respective images. Higher \snr values are required for \cbc{s} with lower chirp masses to maintain equal degree of signal visibility.}
        \label{ch:RDL:fig:A_intr_mc_gamma}
    \end{figure*}

    We, therefore, obtain a metric that is a direct function of the visibility of the signal features in the input image and is independent of the component masses of the \cbc. As an example, the injections shown in Figure~\ref{ch:RDL:fig:visibility_difference} having nearly the same \snr values are fairly separated in their $\gamma$ values which come out to be $26.5$ and $44.9$, respectively. The non-inclusion of the spin parameters in the $A_\mathrm{intr}$ function is expected to retain some impurity in the training data. However, these impurities are expected to be small enough that their effect on training remains minimal. In Figure~\ref{ch:RDL:fig:A_intr_mc_gamma}, we show how the visible strength of signals in the image changes as a function of $\gamma$ alone while the chirp mass only changes the signal morphology, leaving the visible strength of the signal almost unaffected. The recovered \snr{s} are annotated on the respective images for comparison. The \snr values can be seen to increase with reducing chirp mass for a given $\gamma$ bin.

\subsection{Preparation of datasets}\label{ch:RDL:sec:dbprep}

    For generating the \inj dataset, we consider a region of \cbc parameter space covering masses that are most likely to be observed with LIGO/Virgo detectors, and precession. A large number of injections are made in the simulated coloured Gaussian noise generated from \texttt{aLIGOZeroDetHighPower} \psd \footnote{Note that in Section~\ref{ch:RDL:sec:search_real_data}, we perform a search over the LIGO data from the third observing run. The \emph{shape} of the \psd for this data is approximately the same as that of \texttt{aLIGOZeroDetHighPower} \psd, except in the lower frequencies $< 30$ Hz~\cite{Abbott2018}. Since we are dealing with whitened strain data, the \psd shape affects the intensity of the \cbc signals in various frequency bands of \cwt maps, but variations in the overall noise floor do not have any impact. As the \psd values at the lower frequencies are significantly higher than those in the sensitive band of $80-1000$ Hz, the low-frequency variation in \psd is expected to only affect the visibility of the low-frequency tail for the loud \cbc signals in the training data, where the classifier performance is usually reliable.}. The hyper-parameter choices for these injections are listed in Table~\ref{ch:RDL:tab:inj_config}. We whiten and band-pass the data between $10-512$ Hz and collect $3$ second long slices containing injections. Further, we obtain the time-frequency representations of these data slices using \cwt with Morlet wavelet. We then crop these $2$-D arrays to $1$ second long \cwt maps in such a way that the merger times of the binaries are uniformly distributed between $0.7-0.85$ second of the slice. Such a range is chosen to allow a reasonable coverage of the inspiral and post-merger parts of the low-mass and high-mass binaries, respectively. To discard the filter artefacts, we manually set the values corresponding to frequencies less than $16$ Hz in the \cwt maps to zero. For better visualisation, the frequency scale is set to logarithmic. The same procedure is repeated for generating the \cleaninj dataset using the same injection set, except without noise this time. The \gn dataset is prepared by collecting the Gaussian noise realisations from the \inj dataset while omitting the \cbc injections. Therefore, at the level of time-series data, the \inj dataset is effectively the direct addition of data slices from \cleaninj and \gn datasets. However, since the image pixels of \cwt maps in our data correspond to the absolute amplitude at the respective locations, this relation is not followed exactly. Nevertheless, such data preparation greatly reduces the possibility of over-fitting the training data, as the features corresponding to noise in \inj and \gn datasets are very close to each other. This forces the model to focus on the uncommon features that correspond to the \cbc signal exclusively. To prepare the \gl dataset, we obtain the class and GPS time information from Gravity Spy~\cite{Zevin2017, Bahaadini2018, Glanzer2022, Soni2021}, cluster the GPS times within $0.1$ second, and choose the data slices such that the glitches are uniformly distributed between $0.25-0.75$ second of the slice. Since no specific morphology is to be captured in the case of glitches, this choice of range is made while ensuring complete coverage. The same procedure mentioned above is repeated to prepare the \cwt maps. For the datasets involving \cbc injections (\inj and \cleaninj), we consider samples with $\gamma > 30$ and $\rho > 5$. Note that we use the \emph{recovered} \snr instead of \emph{injected} \snr while calculating the $\gamma$ values to account for variations due to noise realisation, thereby reducing the error in the estimations of $\gamma$. This is especially helpful in maintaining faithfulness of the injection data even at lower values of $\gamma$. A total of $50000$ samples were collected with these cuts, of which $95\%$ were used for training and the rest for validation.

    We also prepare three separate test datasets using real LIGO data. For the signal class (\injtest), we make injections with the same parameter choices described in Table~\ref{ch:RDL:tab:inj_config}. However, unlike the \inj dataset, these were made in the real data from the entire third observing run of LIGO detectors at Hanford (H1) and Livingston (L1), and with a new seed value to ensure new samples of parameters. For the noise class, we consider two cases of direct relevance for actual searches: the glitches (\gltest) and the background triggers collected through matched filtering (\trigtest). The specification of \gltest dataset is kept the same as that of \gl while ensuring no overlap between the two. For \trigtest dataset, random samples were obtained from $\sim 1.3$ billion triggers collected in each detector by the focused \pycbc \bbh search of $\sim 7.24$ days long chunk of data between May $29$ to June $5$, $2019$ UTC from GWTC-2.1~\cite{gwtc2.1}. As the \cbc events are sporadic, we do not expect any real events to be present in this dataset. We collect $\sim 5000$ samples for each of the datasets.

    \begin{table}
        \setlength{\tabcolsep}{12pt}
        \centering
        \begin{tabular}{l l}
            \hline
            Parameter & Value \\
            \hline
            Chirp-distance & Uniform in ($5$, $400$) Mpc \\
            Component masses & $M_{chirp}$ uniform in ($5$, $60$) $\msun$ \\
            ~ & $q$ uniform in ($1$, $4$) \\
            Component spins & $|S_{1}|,|S_{2}|$ uniform in ($0$, $0.998$) \\
            ~ & randomly orientated over sphere \\
            Approximant & \texttt{IMRPhenomXPHM} \\
            \hline
        \end{tabular}
        \caption{Choices of hyper-parameters for injections used in \inj and \injtest datasets. Additional cutoffs of $\gamma > 30$ and $\rho > 5$ are applied to ensure uniform visible strength of chirp features across the parameter space.}
        \label{ch:RDL:tab:inj_config}
    \end{table}

\section{Where is the model looking?}\label{ch:RDL:sec:looking}

    One of the primary hurdles that limit the reliability of \dl methods is their tendency to over-fit the data in undesirable ways. Here, we explore an approach that significantly mitigates this issue and helps our model focus only on the relevant features in the input data. Also, the signal-specific information is minimal compared to the actual amount of data fed to the model. Therefore, a reduced representation of the input data that captures the complete signal information is particularly beneficial for any analysis. Specifically, for the case of modelled signals, this reduced parameter space can, in principle, be written in terms of only the non-degenerate set of intrinsic and extrinsic parameters of the signals. Therefore, a model that reads the complete signal information from the data should ideally be able to represent the signal in a reduced parameter space that can be transformed back to the full signal without any loss\footnote{Here, the assumption is that the input data representation is obtained using a transformation of the strain data that is lossless for the signal.}. However, the presence of additive noise inevitably introduces statistical errors in estimating source parameters in any space. Though the resulting errors in the inferred parameters can be theoretically characterised using the statistical noise model.

    For the signal model, an extensive matched-filter-based search uses \mle analytically over the extrinsic parameters and numerically over the intrinsic parameters. \mle essentially achieves the reduced representation of the input strain in the parameter space of \cbc signals. However, such searches are computationally expensive, which limits the coverage of some parameters, such as eccentricity, component spins, etc. Full coverage of such parameters could increase the computation cost by orders of magnitude~\cite{PhysRevD.94.024012,Nitz_2020}, which is presently impossible to accommodate. Also, the noise model considered for constructing the log-likelihood function for the search assumes stationary Gaussian noise for calculating the signal-to-noise ratio~\cite{creighton_anderson}. In general, search models built with this assumption (matched-filter-based searches or \dl approaches trained against Gaussian noise alone) prove inadequate when the noise has non-Gaussianities in the form of glitches. Including glitches in the construction of the search log-likelihood function can play a crucial role in separating signal and noise more effectively. However, this is not achievable in practice due to the lack of an analytical description of the glitches. Rather, various signal consistency tests are often employed as additional measures to mitigate the adverse effect of glitches on search sensitivity~\cite{Allen_chi,Nitz2018}. \dl models being excellent approximators of non-linear features can learn the broad set of morphologies followed by these glitches. In the presence of Gaussian noise alone, the performance of a \dl model can be compared against that of matched filtering. However, their added ability to learn artefacts originating from glitches allows them to surpass matched-filter-based searches in principle. However, these arguments are based on theoretical assumptions of the capabilities of \dl models and achieving such performance in practice is often very challenging. 

\subsection{Introduction to VAEs}\label{ch:RDL:sec:vae_intro}

    A specific class of neural networks, known as auto-encoders, offer encoding of the data into an efficient coding~\cite{Goodfellow_DL}. The architecture of auto-encoders involves an encoder $\mathcal{E}_\phi$ that encodes the input data $\mathcal{X}$ into a latent space coding $\mathcal{Z}$ and a decoder $\D$ that learns to regenerate the required output data $\mathcal{X'}$ from it, where $\mathcal{X'}$ can be same as $\mathcal{X}$ or a transformed version of $\mathcal{X}$. The reduced coding $\mathcal{Z}$ offers a condensed representation of the data. Going a step beyond, a \vae offers the representation of each input sample $x$ as a multivariate posterior distribution $q_\phi(z|x)$ for $x \in \mathcal{X}$ and $z \in \mathcal{Z}$~\cite{kingma_VAE}. Typically, the posterior $q_\phi(z|x)$ is modelled as a Gaussian and is regularised to be as close as possible to the prior of $\D$ using the \kl divergence. This prior is denoted by $p_\mathrm{model}(z)$ and is employed for generating new data using $\D$. It is generally taken as the Normal distribution $\mathcal{N}(0,1)$. As values of $z$ are drawn from the distribution $q_\phi(z|x)$ at the latent layer, performing back-propagation becomes problematic. This issue is solved using the re-parametrisation trick, which models the distribution $q_\phi(z|x)$ as a multivariate Gaussian whose mean, $\mu$ and standard deviation, $\sigma$ are predicted as latent space outputs from the encoder. Now, $q_\phi(z|x)$ can be written as, 
    \begin{equation}
        q_\phi(z|x) = \mu + \mbox{\boldmath$\epsilon$} \cdot \sigma,
    \end{equation}
    where the random variable {\boldmath$\epsilon$} is described by the independent fixed distribution $\mathcal{N}(0,1)$. Therefore, the back-propagation is made possible with the gradient descent performed over the parameters $\mu$ and $\sigma$ that govern the distribution $q_\phi(z|x)$ instead of the sampled values $z$. The final loss function to train a \vae, given by Equation~\ref{ch:RDL:eq:vae_loss}, has two components. First, the reconstruction loss that compares the output against $\mathcal{X'}$, and second, the \kl divergence-based loss. The loss terms are decomposed into the sum of losses coming from each sample $j$ from the training data.

    \begin{widetext}
    \begin{equation}\label{ch:RDL:eq:vae_loss}
        \begin{split}
            \mathcal{L}_{\phi,\theta}(\mathcal{X}, \mathcal{X'}, q) 
                &= \mathcal{L}_\mathrm{rec}(\mathcal{X}, \mathcal{X'}) + \sum_j \mathcal{D}_\mathrm{KL}\big(q_\phi(z_j|x_j) \parallel \mathcal{N}(0,1)\big)\\
                &= \sum_j \Big[ \big\|x'_j - \mathcal{D}_\theta(\mathcal{E}_\phi(x_j))\big\|^2 + \mathcal{D}_\mathrm{KL}(\mathcal{N}(\mu_j,\sigma_j) \parallel \mathcal{N}(0,1)) \Big].
        \end{split}
    \end{equation}
    \end{widetext}

    Training of the \vae model with such a loss function instils bounded, smooth and meaningful parametrisation of the latent code $\mathcal{Z}$. 

    \begin{figure}
        \centering
        \includegraphics[width=\linewidth]{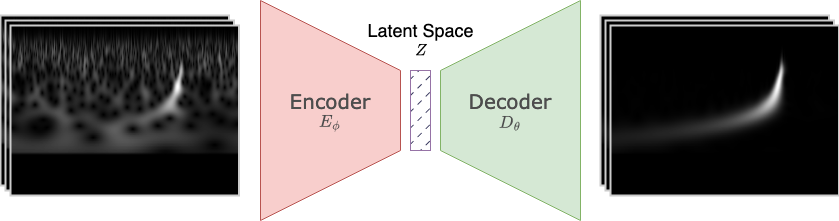}
        \caption{The schematic of the VAE model, along with its input (\inj) and target output (\cleaninj) used for training. The sampling layer at the latent space learns to represent the \cbc features extracted from the input as a multivariate Gaussian, thereby obtaining their reduced and smooth representation.}
        \label{ch:RDL:fig:vae_arch_used}
    \end{figure}

    \begin{figure}
        \centering
        \includegraphics[clip=true, trim=1.5cm 0cm 2cm 2cm, width=\linewidth]{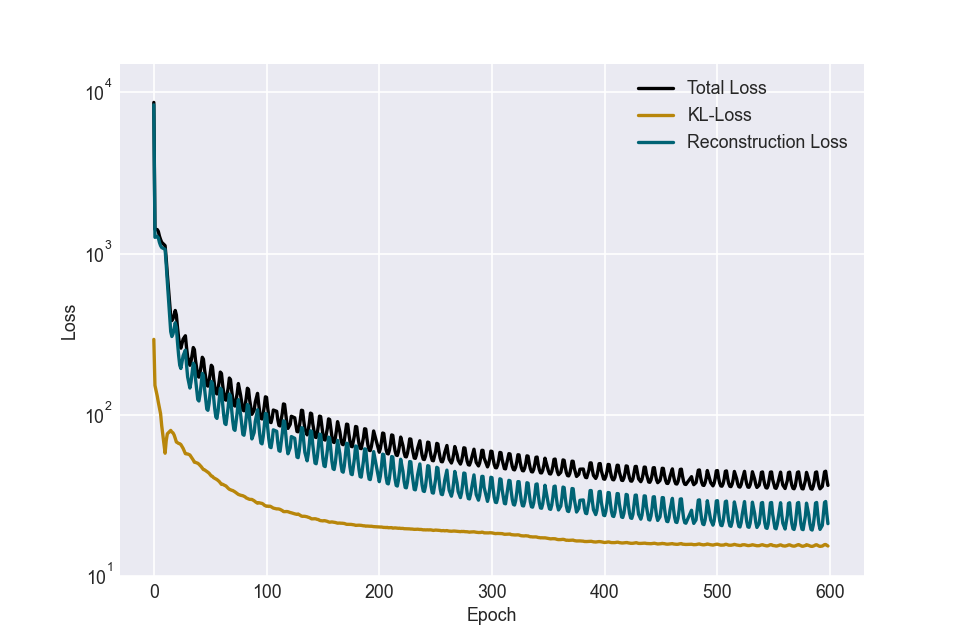}
        \caption{Evolution of losses over $600$ epochs of training. The \kl divergence saturates to an almost constant value after $\sim 400$ epochs while the reconstruction loss continues to reduce slowly. Note that the y-axis is in log scale. The \vae model reaches a satisfactory level of reconstruction ability at the end of training.
        }
        \label{ch:RDL:fig:vae_losses}
    \end{figure}

    \begin{figure}
        \centering
        \includegraphics[clip=true, width=\linewidth]{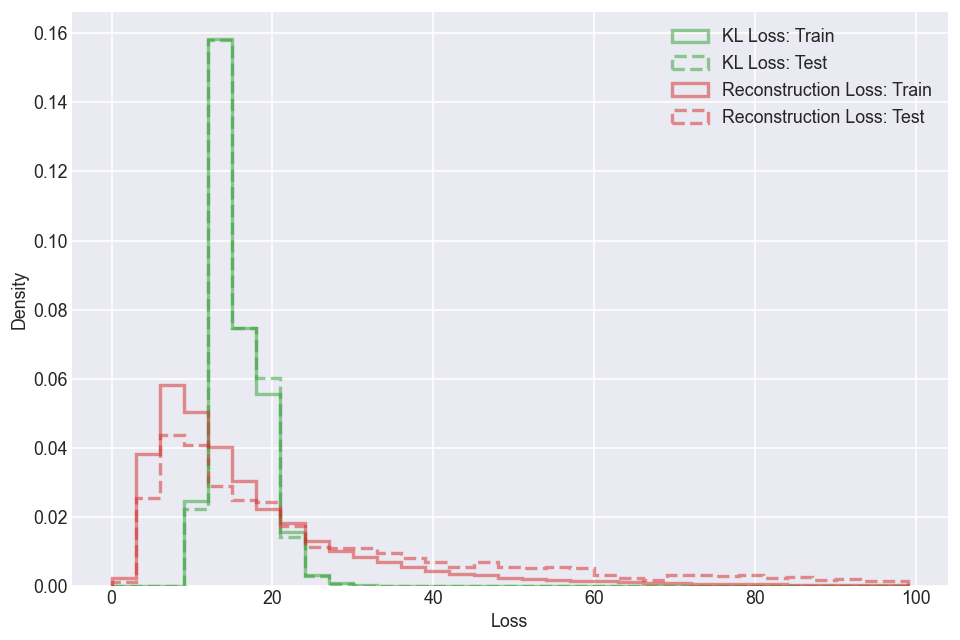}
        \caption{Histograms of \kl loss (green) and reconstruction loss (red) for the \vae on training data (solid line) and test data (dashed line). The histogram for reconstruction loss shows mild overfitting to the training data. This is a commonly observed phenomenon during the training of \dl models resulting in the difference between the training and validation performance.
        }
        \label{ch:RDL:fig:vae_hist}
    \end{figure}

\subsection{Reduced representation of signal space using VAE}\label{ch:RDL:sec:vae_impl}

\begin{figure*}
        \subfloat[Structure of the parallel module.]
        {%
            \centering
            \hfill
            \includegraphics[width=0.45\textwidth]{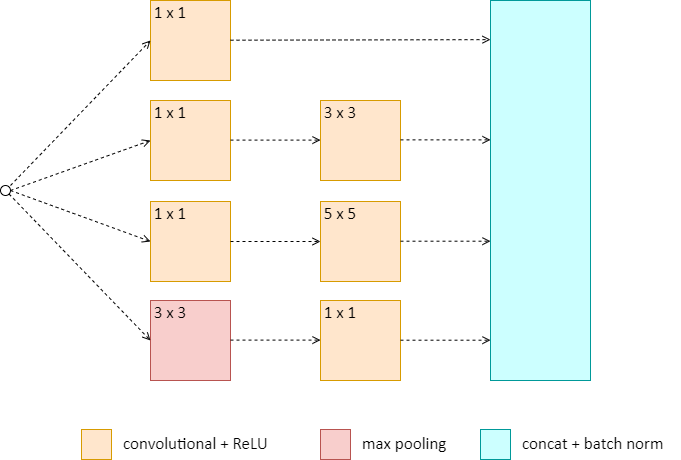}%
            \vspace{0.3cm}
        } \hfill
        \subfloat[Sampling layer at the latent space.]
        {%
            \includegraphics[width=0.45\textwidth]{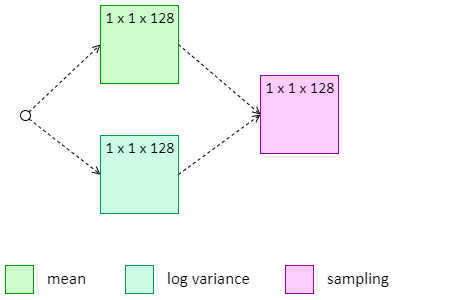}%
            \hfill
            \vspace{0.3cm}
        } \vspace{0.5cm}
        \subfloat[Overall encoder architecture.]
        {%
            \includegraphics[width=0.9\textwidth]{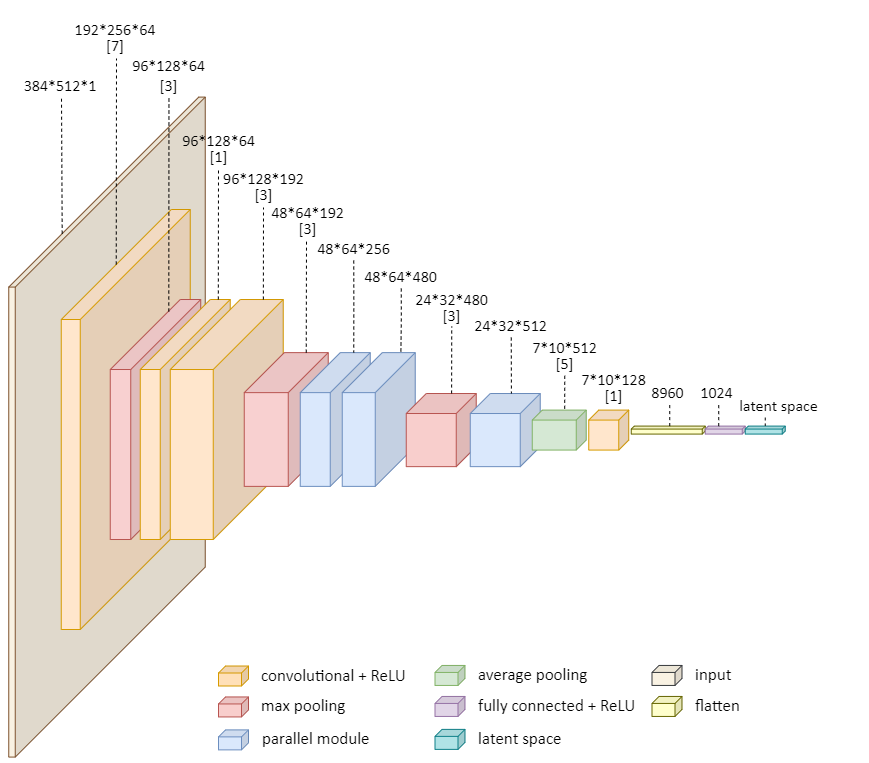}%
            \vspace{0.3cm}
        }
        \caption{Architecture of the encoder part of the \vae model along with its sub-components. Outputs of different layers, colour coded for distinction, are shown. The shapes of these outputs are given at the top with the number in the bracket denoting the size of the square kernel. The encoder consists of three parallel modules, which have convolutions in parallel and in series, inspired by the inception architecture~\cite{Szegedy2016}. The latent space has a sampling layer for the multi-variate Gaussian distribution represented by the mean and variance vectors.
        }
        \label{ch:RDL:fig:arch_vae_encoder}
    \end{figure*}

\begin{figure*}
        \subfloat[Branch structure.]
        {%
            \hspace{2cm}
            \includegraphics[width=0.35\textwidth]{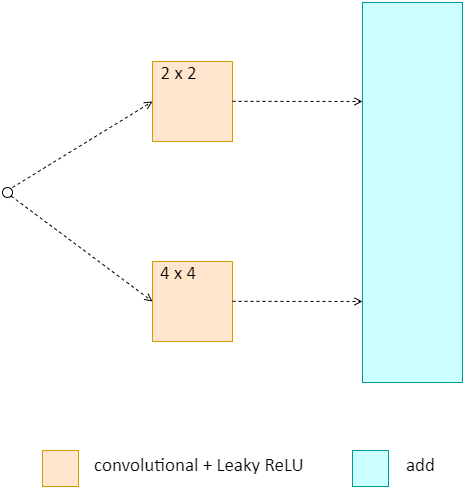}
            \vspace{0.3cm}
        } \hfill \vspace{0.5cm}
        \subfloat[Overall decoder architecture.]
        {%
            \centering
            \includegraphics[width=0.9\textwidth]{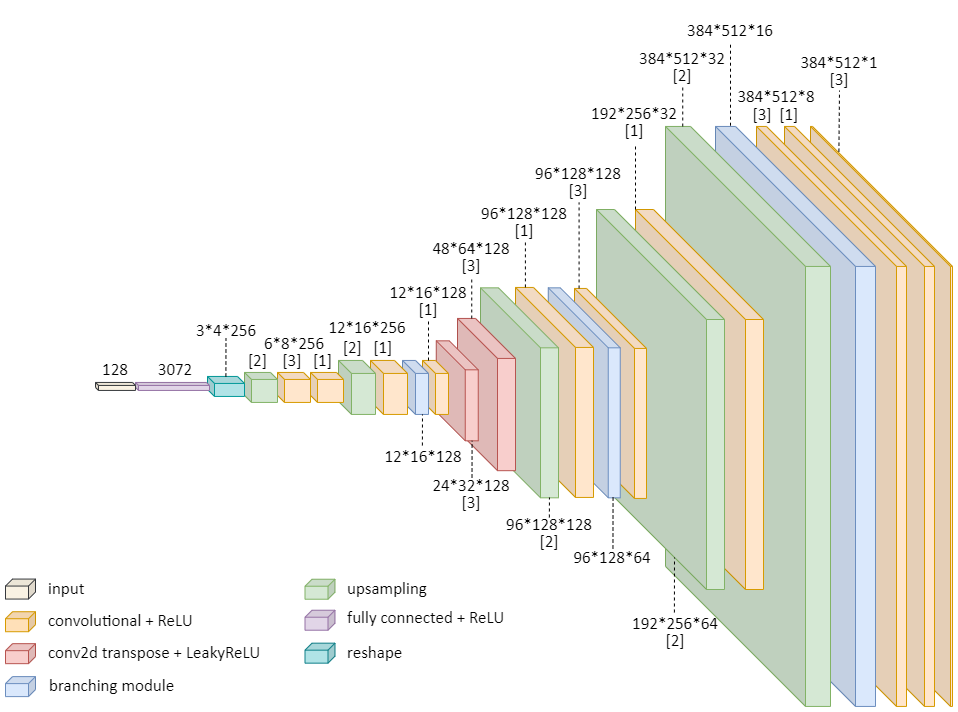}
            \vspace{0.3cm}
        }
        \caption{Architecture of the decoder part of the \vae model along with its branch structure. The depictions are similar to those shown in Figure~\ref{ch:RDL:fig:arch_vae_encoder} except for the difference in the types of layers. The inputs to the decoder are the latent space mean vector codings of the \inj dataset and the outputs are the corresponding images from the \cleaninj dataset.
        }
        \label{ch:RDL:fig:arch_vae_decoder}
    \end{figure*}

    We build a \vae for the task of \emph{denoising} the \cwt maps. However, the denoising task here is not the literal pixel-level denoising of the image. Instead, it is transforming the input \cwt maps from \inj into the corresponding \cwt maps of \cleaninj, which demands preserving of features belonging to the injection exclusively and ignoring other artefacts belonging to the noise class as shown in Figure~\ref{ch:RDL:fig:vae_arch_used}. Therefore, during training, the model learns to \emph{look for} only the signal features, encode them into the latent space and decode a pure estimate of the signal. Since our final classification task is only signal-specific, any lack of retention of other non-signal-specific information from the input, such as glitch or Gaussian noise features, is not a concern. Rather, the insensitivity of the encoder to some of the glitch features may be desirable for eliminating the respective glitches right at the first step of inference. We construct the encoder part of the \vae model using sub-modules consisting of parallel branches of convolutions of different kernel sizes as shown in Figure~\ref{ch:RDL:fig:arch_vae_encoder}. The parallel modules are inspired by Inception modules which help extract features at varying scales~\cite{Szegedy2016}. Similarly, the architecture of the decoder part is shown in Figure~\ref{ch:RDL:fig:arch_vae_decoder}. We use L$1$ regularisation with $l_1 = 0.001$ for kernel and bias parameters in all convolution layers of the encoder and decoder. The weights of convolutional kernels were initialised using Xavier uniform initialiser~\cite{Glorot2010} while biases were initialised with a constant value of $0.2$. We use a cyclical learning rate varying in the range $10^{-5}-10^{-4}$, which resulted in faster convergence and better performance at the same time~\cite{Smith2015}. Since, such a scheme results in oscillating losses, we use a callback function to monitor the total loss during training and store the best checkpoint with minimum loss. The choice of optimiser was stochastic gradient descent (SGD) with momentum $0.01$. We also set the gradient clipping at $0.5$ to deal with the exploding gradient problem. We trained our \vae model on four NVIDIA A$100$ Tensor Core \glspl{gpu} for $600$ epochs (run-time $\sim 11$ hours).

\begin{figure}
        \centering
        \includegraphics[width=\linewidth]{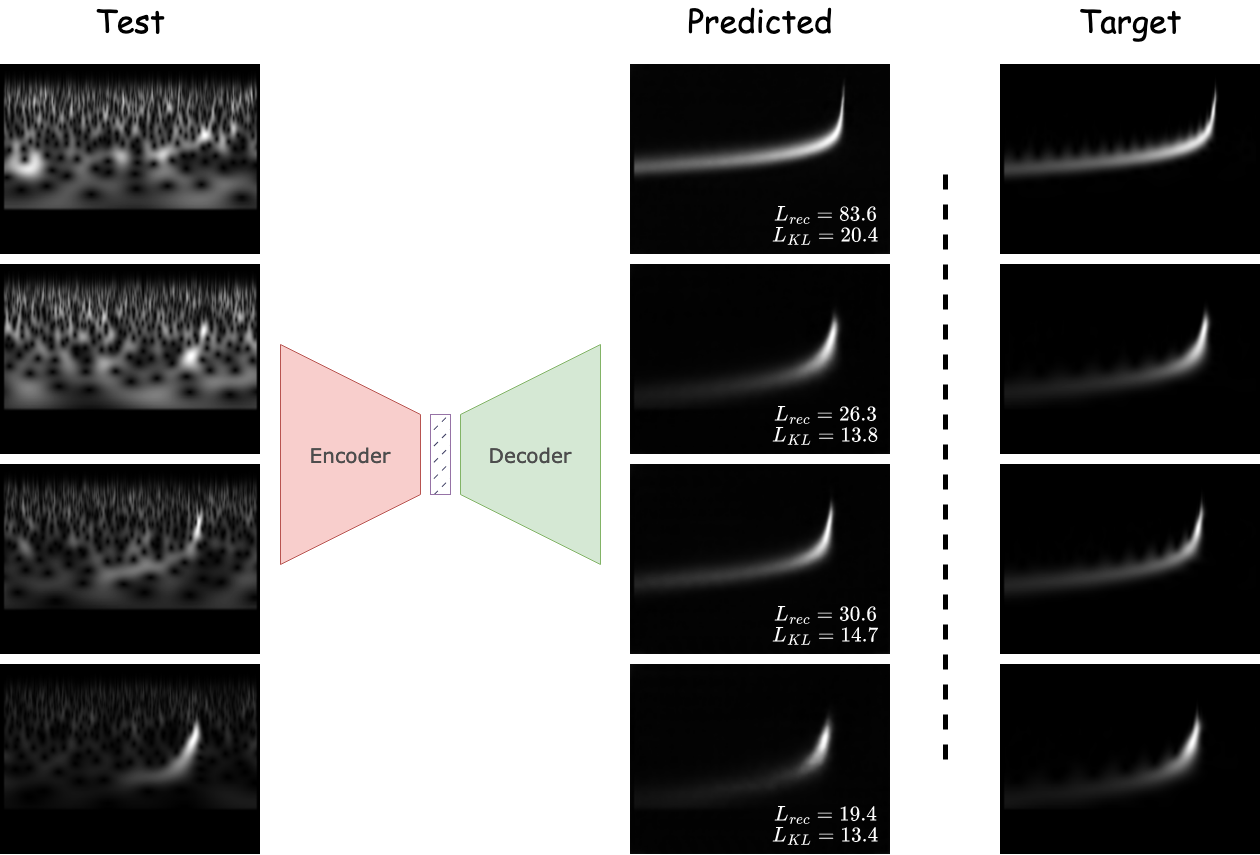}
        \caption{Reconstructed clean maps output by the trained \vae model along with their respective reconstruction loss and \kl loss values are shown for some of the input samples from the validation \inj dataset. The corresponding target samples from the \cleaninj dataset are also shown for comparison.
        }
        \label{ch:RDL:fig:vae_reconstructed_figs}
    \end{figure}

    Figure~\ref{ch:RDL:fig:vae_losses} shows the evolution of losses during model training with an overall asymptotically saturating behaviour with oscillations originating from cyclical learning. A comparison of the performance of the trained \vae model on training and validation \inj datasets is shown in Figure~\ref{ch:RDL:fig:vae_hist}. The histograms of \kl loss for training and validation datasets are similar. Whereas, the reconstruction loss exhibits slight overfitting towards training data, a tendency that is commonly observed in the training of \dl models and is not a cause of serious concern unless the predictions for training and validation data show significant variations from each other. The best-collected checkpoint had a reconstruction loss value of $19.3$ and the \kl divergence loss value of $15.3$. To showcase the model's ability to reconstruct clean signals from input injections, in Figure~\ref{ch:RDL:fig:vae_reconstructed_figs}, we show $4$ randomly selected images from the validation \inj data, their reconstructions using the trained \vae model and the target \cleaninj images for comparison. The reconstructed images are annotated with corresponding losses for reference. Once trained, we discard the decoder part of the model and only use the encoder, which gives the reduced latent code of the input image as an output.

    \begin{figure*}[htb]
        \subfloat[\label{ch:RDL:fig:tsne_vanilla_model}]
        {%
            \centering
            \includegraphics[clip=true, trim=3cm 3cm 2.4cm 3cm, width=0.48\linewidth]{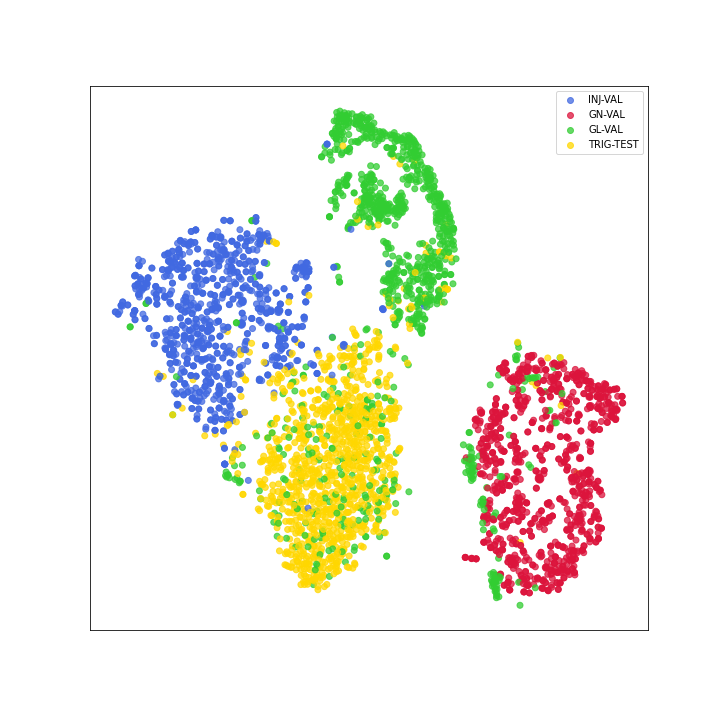}
        }\hfill
        \subfloat[\label{ch:RDL:fig:tsne_fine-tuned_model}]
        {%
            \centering
            \includegraphics[clip=true, trim=3cm 3cm 2.4cm 3cm, width=0.48\linewidth]{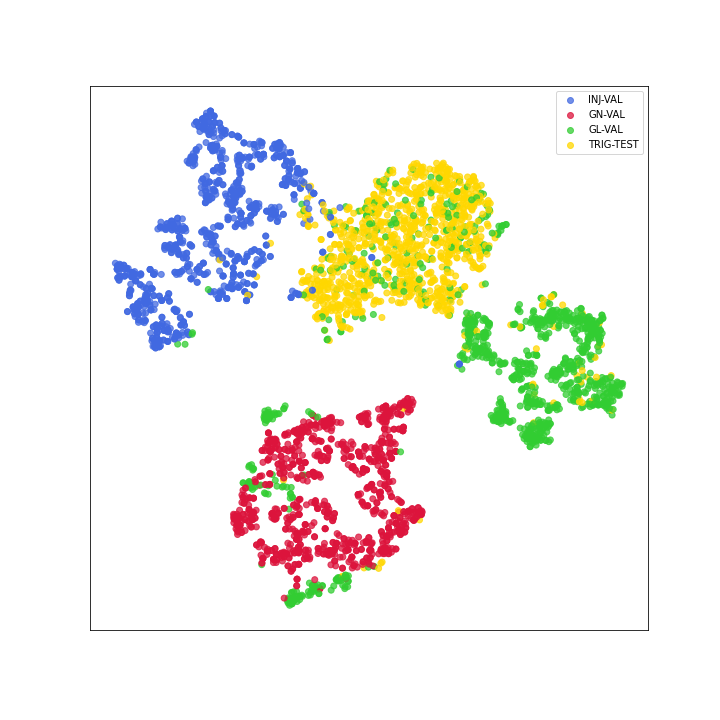}
        }
        \caption{Latent space representations of the validation data from \inj, \gn and \gl datasets along with the \trigtest dataset visualised using t-SNE at the encoder output. (a) Encoder output taken from the trained \vae model. Though the \vae was trained on \inj and \cleaninj data alone, the encoder can already separate the classes very well. (b) Encoder output from the final classifier after fine-tuning. Since the classifier was trained on \inj, \gn and \gl datasets, an improved class separation can be observed after fine-tuning.}
        \label{ch:RDL:fig:tsne_visualisations}
    \end{figure*}

    We use \tsne to visualise the latent space encoding at different stages of model development and make qualitative observations on their performance. \tsne is a statistical method for visualising high-dimensional data by giving each data point a location in a two or three-dimensional map~\cite{tSNE}. Unlike \pca, a linear dimensionality reduction algorithm, \tsne is based on probability distributions with a random walk on neighbourhood graphs to interpret the complex polynomial relationships between features and find structure within the data. We generate the \tsne visualisations of the latent space coding obtained from the encoder at two separate stages. The first being the encoder output from the trained \vae model, while the latter one is obtained after the classifier fine-tuning is accomplished (described in Section~\ref{ch:RDL:sec:utility}). The prime difference between the two is that the first model was trained with the \inj dataset alone, while the second one was trained with \inj, \gn and \gl datasets. To obtain the latent space required for \tsne, we analyse the validation data samples from \inj, \gn, \gl and test data samples from \trigtest with both the encoder models. Before proceeding with \tsne visualisation, we reduce the dimensions of the latent space from $256$ to $10$ using \pca. The benefits are two-fold, it helps suppress some noise in the final representation and speeds up the algorithm without losing much information. After performing \pca, we find that the explained variance ratio for the encoder output after \vae training was $0.46$ compared to $0.94$ after the fine-tuning stage. From the \tsne representations shown in Figure~\ref{ch:RDL:fig:tsne_visualisations}, it is evident that even without training on the \gn, \gl and \trigtest datasets, the initial encoder is already able to separate them satisfactorily. The class separation improves further after fine-tuning.

\section{Utility of the latent code for classification}\label{ch:RDL:sec:utility}

    \begin{figure}
        \centering
        \includegraphics[width=0.9\linewidth]{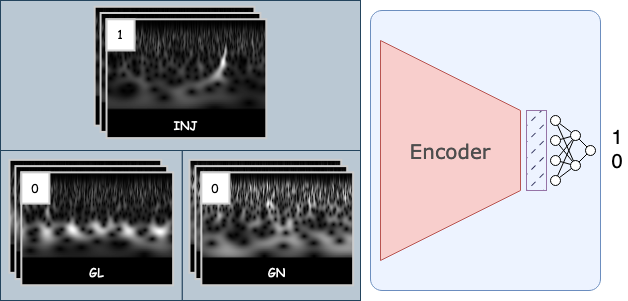}
        \caption{A hybrid classifier model obtained by appending densely connected layers on top of the trained \vae's encoder. The network is trained to map the latent space coding to the binary classification output corresponding to signal (\inj) and noise (\gl and \gn with losses weighed down by half).
        }
        \label{ch:RDL:fig:hybridNN_arch}
    \end{figure}

    \begin{figure*}[htb]
        \subfloat[]
        {%
            \centering
            \includegraphics[clip=true, trim=1.5cm 1cm 2cm 2cm, width=0.48\linewidth]{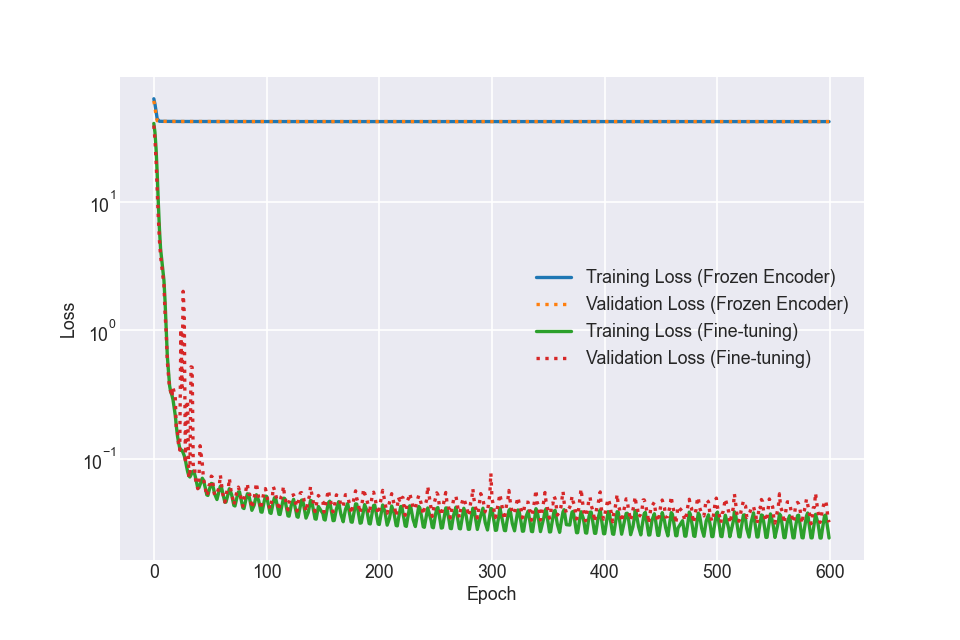}
        }\hfill
        \subfloat[]
        {%
            \centering
            \includegraphics[clip=true, trim=1.5cm 1cm 2cm 2cm, width=0.48\linewidth]{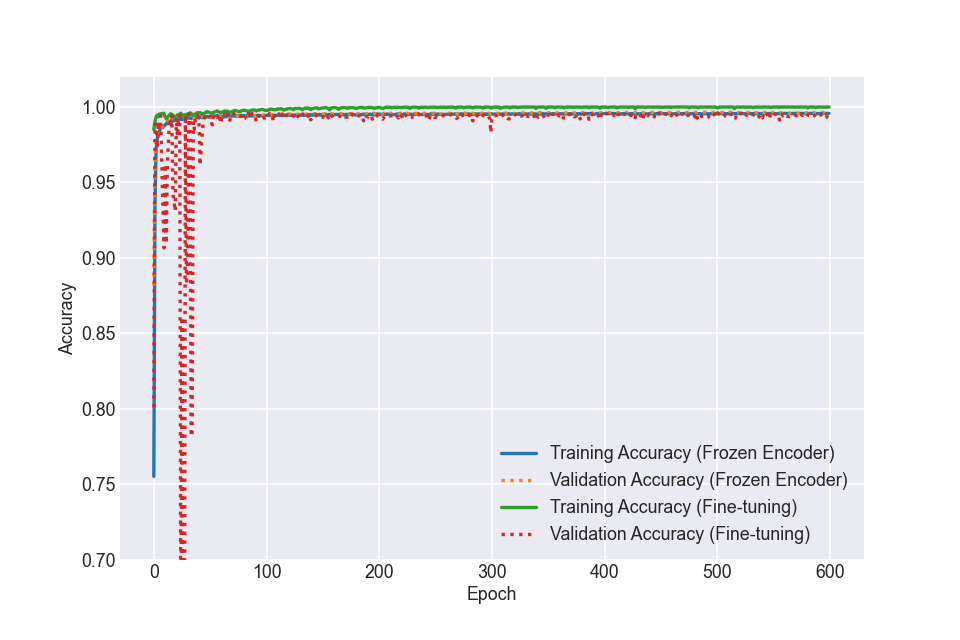}
        }
        \caption{Performance metrics for the hybrid classifier are shown for its two stages of training: with encoder frozen and fine-tuning of the entire model. During both stages, the classifier is trained to separate the \inj class from the noise classes (\gn and \gl). (a) Evolution of binary cross-entropy losses for training and validation data. Initially, with the encoder frozen, both training and validation losses saturated to a value of $42.4$. During fine-tuning, the best checkpoint was collected with a training loss of $0.025$ and a validation loss of $0.031$. (b) Evolution of training and validation accuracies. At the end of the training, both the accuracies for the hybrid model with frozen encoder were close to $99.6\%$. The best checkpoint had a training accuracy of $99.99\%$ and a validation accuracy of $99.55\%$.}
        \label{ch:RDL:fig:hybridNN_training}
    \end{figure*}

    We develop a hybrid classifier model by appending the encoder with additional densely connected layers that map the latent space to a single output node predicting the presence of a \cbc signal in the input (refer to Figure~\ref{ch:RDL:fig:hybridNN_arch}). The input to the densely connected layers is the mean-vector {\boldmath$\mu$} of the multivariate Gaussian distribution from the latent space. We train this hybrid model in two stages. Initially, we froze the encoder and only allowed the appended dense layers to learn the mapping between the latent space and the binary outputs. The training data was composed of signal class (\inj dataset) and noise class (\gn and \gl datasets) with $1$'s and $0$'s as targets, respectively. The losses for \gn and \gl datasets were weighed down by half to maintain class balance. The data were shuffled before the start of training. We train the hybrid network for $600$ epochs (run-time $\sim 5$ hours $30$ minutes) with a cyclical learning rate varying between $10^{-5}-10^{-4}$ and binary cross-entropy as the loss function with label smoothing set to $0.005$. We used Adam optimiser with $\beta_1 = 0.9$, $\beta_2 = 0.999$ and $\epsilon = 10^{-7}$. The gradients were clipped at a maximum value of $50$. At the end of the training of the first stage, both training and validation accuracies were close to $99.6\%$ while the losses saturated around $42.4$. After the parameters of the densely connected layers were learnt, we further fine-tuned the network for $600$ epochs (run-time $\sim 6$ hours) by making the entire model, including the encoder, trainable. We used a callback function that monitored the validation loss during training and stored the best checkpoint version of the network. The best checkpoint had a training accuracy of $99.99\%$ and a validation accuracy of $99.55\%$ while the corresponding loss values are $0.025$ and $0.031$, respectively. The evolution of loss and accuracy for the training of the hybrid classifier for both frozen and fine-tuning stages is shown in Figure~\ref{ch:RDL:fig:hybridNN_training}. Note that, especially during fine-tuning, the accuracies appear to stagnate while the binary cross-entropy losses continue to decrease. This indicates an increasing separation between the output distributions corresponding to \inj and noise classes.

\section{Bringing robustness to the model}\label{ch:RDL:sec:robustness}

    \begin{figure}
        \centering
        \begin{tabular}{l}
            \includegraphics[width=0.4\linewidth]{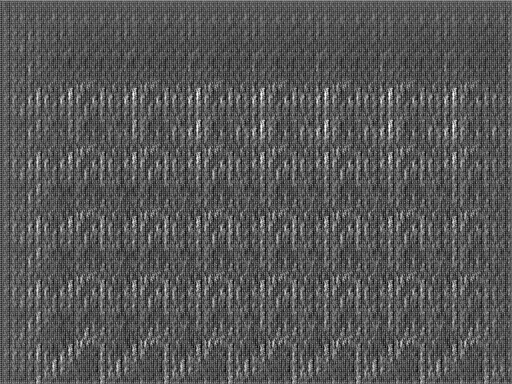} \hspace{0.1cm} \includegraphics[width=0.4\linewidth]{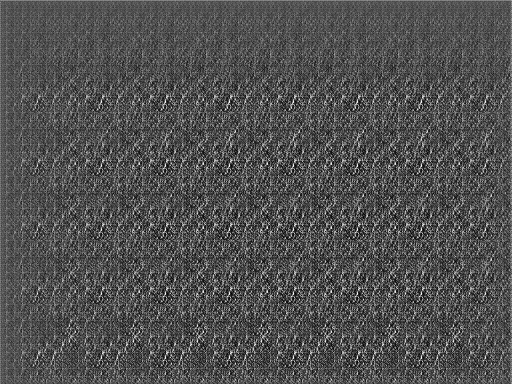} \\[0.1cm]
            \includegraphics[width=0.4\linewidth]{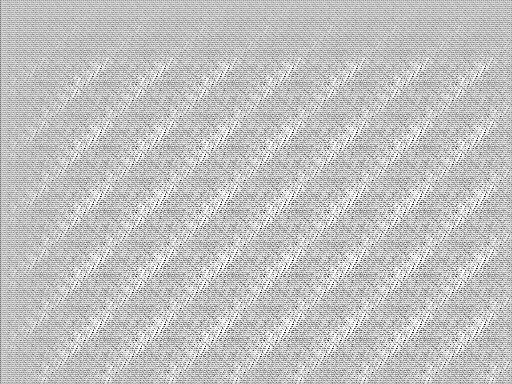} \hspace{0.1cm} \includegraphics[width=0.4\linewidth]{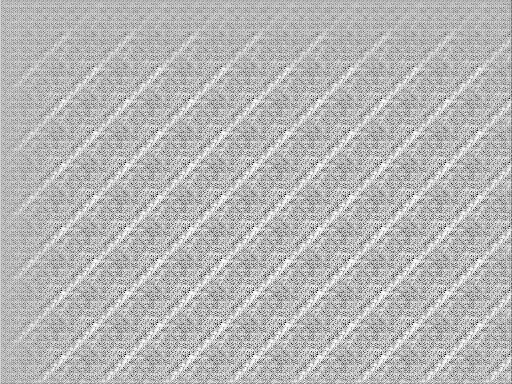}
        \end{tabular}
        \caption{Examples generated by performing adversarial attacks on the fine-tuned classifier model. Despite having no discernible \cbc features, the classifier identified all of these examples as \cbc signals with $\mathrm{P}_\mathrm{cbc}>99\%$.
        }
        \label{ch:RDL:fig:failmodes}
    \end{figure}

    Large \dl models typically show an over-sensitivity to certain features over others~\cite{szegedy2013intriguing}. A class of techniques called \emph{adversarial attacks} exploits this weakness of \dl models and finds their failure modes. This can either be accomplished by systematically perturbing the input data with \emph{gradient ascent} or training a separate generative adversarial \dl model. The first kind, that of perturbative attack on the regular input, can be as simple as a single pixel attack~\cite{singlepixattack} or a more generic attack~\cite{goodfellow2014explaining}, leading to a dramatically degraded model performance. These perturbations in the data are typically insignificant when perceived by a human. The second kind of attack can be achieved through generative adversarial frameworks which involve training generative \dl models which can create images that deceive the discriminator successfully. Initialised randomly, these models learn to generate images that do not contain any practically meaningful information, yet are able to mislead the classifier~\cite{goodfellow_GAN,Nguyen2014}.

    Our fine-tuned classifier shows very promising performance provided the inference is made on data that belong to the same input space as the training data. However, as mentioned above, the model may show a dramatically degraded performance if the data for inference deviates from this space. Such deviations could occur due to changes in the data characteristics or the pre-processing scheme. Such changes can render the model inapplicable, and retraining the model with an updated dataset may be required each time they occur. This can become especially challenging in certain cases, for example, new types of glitches that may keep arising with changing operating conditions of the detectors.

    In recent years, multiple strategies have been explored to make \dl models more robust~\cite{goodfellow2014explaining,GANdef,RobGAN}. Particularly, including adversarially perturbed images in training has been shown to improve the robustness of the models to small perturbations~\cite{goodfellow2014explaining}. However, it leads to robustness against perturbations only in the vicinity of the input space covered by training data. As we intend to explore the failure modes in a parameter space that is not necessarily close to the input image space, we take a somewhat different approach of using the generative adversarial framework to bring robustness to our classifier, as described in the next section.

\subsection{Introduction to GANs}

    \begin{figure}
        \centering
        \includegraphics[width=\linewidth]{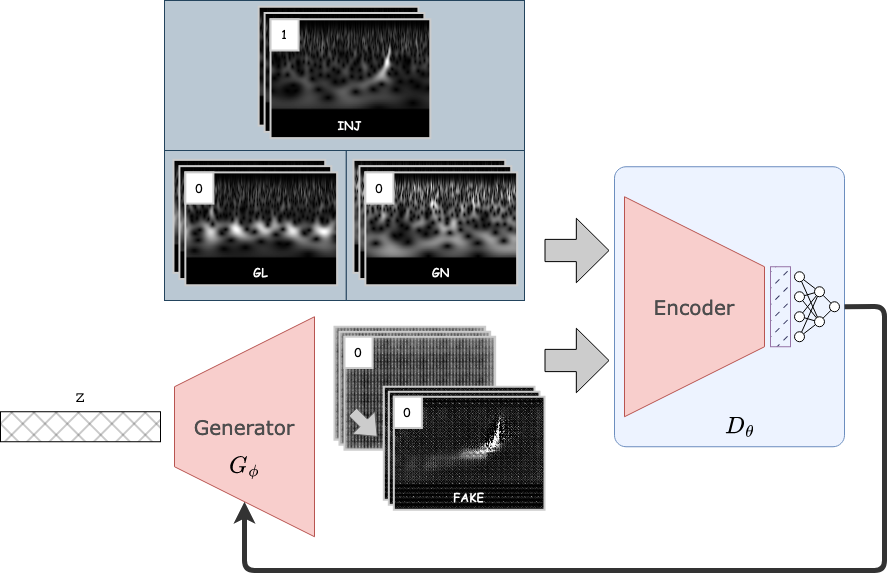}
        \caption{Schematic of the GAN setup with the hybrid classifier being trained as the discriminator alongside $5$ adversarial generator models tasked with finding its failure modes. The training data was the same as that used for training the hybrid classifier earlier. The initially unrealistic fake images produced by the generators eventually converge to images similar to those from the \inj dataset.}
        \label{ch:RDL:fig:gan_arch}
    \end{figure}

    A \gan is composed of two \dl models, a generator $\G$ and a discriminator $\D$, playing a minimax game and being trained in the process~\cite{goodfellow_GAN}. The generator uses a latent space input $z$ to generate fake data $\G(z)$. The discriminator analyses this data and gives the output predictions $\D(\G(z))$ of them being real. In such a setup, $\D$ is trained for correctly classifying the training data $x$ from the fake data $\G(z)$ generated by $\G$. In other words, it is trained with $1$'s and $0$'s as targets for $\D(x)$ and $\D(\G(z))$, respectively. Therefore, for $\D$, the training objective is the minimisation of a loss function given by~\cite{goodfellow_GAN},
    \begin{equation}
    \label{ch:RDL:eq:D_loss}
        \mathcal{L}(\theta) = - \bigg[ \mathrm{log}\D(\overbar{X}) + \mathrm{log}(1-\D(\G(\overbar{Z}))) \bigg],
    \end{equation}
    through gradient descent. Simultaneously, $\G$ is trained to generate fake data that can mislead the discriminator into giving positive outputs. The loss function that is to be minimised in this case is~\cite{goodfellow_GAN},
    \begin{equation}
    \label{ch:RDL:eq:G_loss}
        \mathcal{L}(\phi) = - \mathrm{log}(\D(\G(\overbar{Z}))).
    \end{equation}
    Provided $\G$ and $\D$ have enough capacity, it can be proved that the optimisation algorithm saturates when the generator learns to produce data that is indistinguishable from the real data~\cite{goodfellow_GAN}, i.e., the posterior of $\G$, $p_g$ converges to that of data, $p_\mathrm{data}$ which is the global optimum of the minimax game. At this point, the output of $\D$ saturates to $\D^{*}(x) = \frac{p_\mathrm{data}}{p_\mathrm{data} + p_g}$, which, for equal proportions of training and generated data, is $\frac{1}{2}$.

\subsection{Fragility of classifier and proposal for robustness}\label{ch:RDL:sec:robustness_proposal}

    \begin{figure*}[htb]
        \subfloat[GAN Discriminator Losses]
        {%
            \centering
            \includegraphics[clip=true, trim=1.5cm 1cm 2cm 2cm, width=0.48\linewidth]{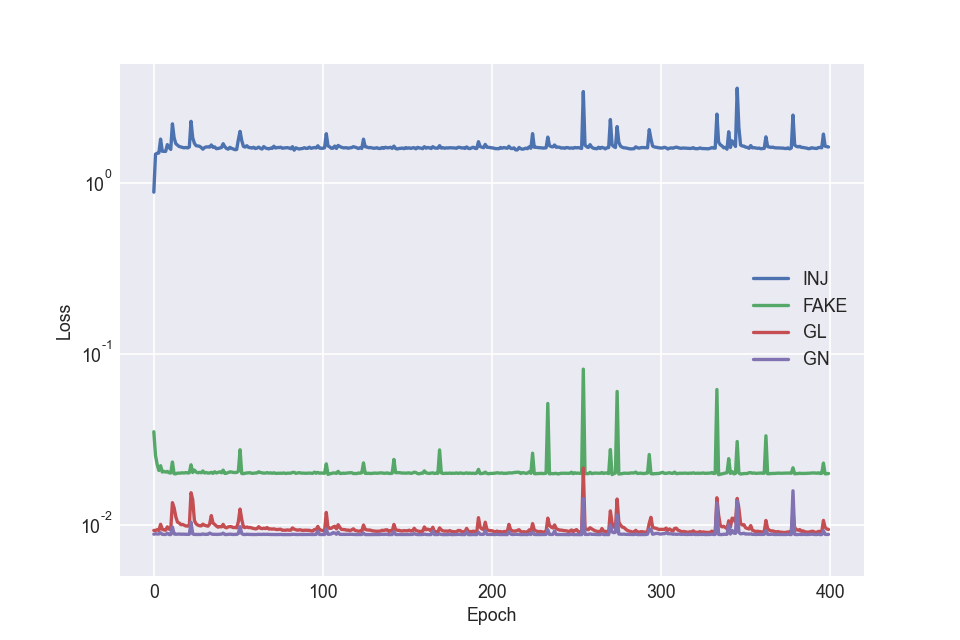}
        }\hfill
        \subfloat[GAN Generator Losses]
        {%
            \centering
            \includegraphics[clip=true, trim=1.5cm 1cm 2cm 2cm, width=0.48\linewidth]{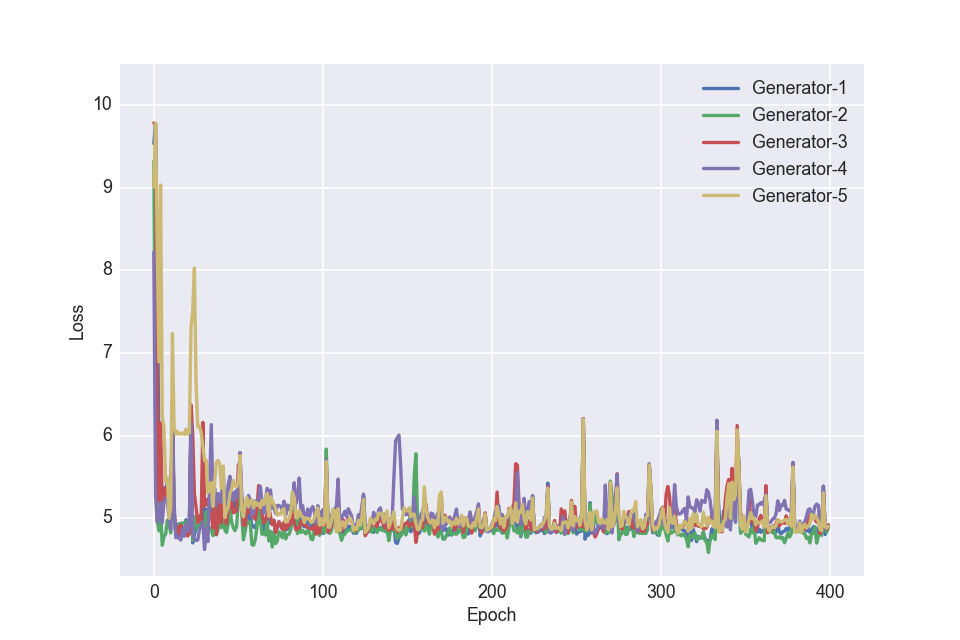}
        }
        \caption{(a) Evolution of discriminator losses given by Equation~\ref{ch:RDL:eq:D_loss} in the \gan setup during training of $400$ epochs. Loss on fake data generated by the generators is shown along with the losses for \inj, \gn and \gl data. (b) Progress of generator losses given in Equation~\ref{ch:RDL:eq:G_loss} through the training. Five generators were trained simultaneously alongside the discriminator. The initial incapability of generators is overcome quickly in the first $\sim 30$ epochs.}
        \label{ch:RDL:fig:gan_losses}
    \end{figure*}

    Firstly, we demonstrate the fragility of our fine-tuned model in its original state by identifying its failure modes. We trained five adversarial generators with the loss function given in Equation~\ref{ch:RDL:eq:G_loss} against the classifier, which was frozen. Training in this setup is equivalent to training only the adversaries in a \gan to find the failure modes of the discriminator. The adversaries were initialised with random weights at the start of training. Hence, their output images are unrealistic in the initial epochs. However, we observed that the adversaries learn quickly, within $\sim 5$ epochs, and produce images classified as signals by the classifier with very high output probabilities. Figure~\ref{ch:RDL:fig:failmodes} shows a few sample images generated by the adversaries at the end of $5^\mathrm{th}$ epoch. It can be seen that these images do not belong to any meaningful category, and yet, are classified as \cbc signals with output values $> 0.99$.

    To train our classifier against its failure modes, we trained the discriminator alongside the five adversaries in the same \gan framework. The motivation for such training is to identify and exhaust all the possible failure modes of the classifier by simultaneously besetting it with several adversaries while the training is still in progress. It would lead to an ultimate saturation point when the adversaries start producing images very close to the signal space in the training data and exhaust all other possibilities.

    However, there are several pitfalls in this proposal. We need to modify the regular \gan framework to achieve desirable performance for the required task of robust classification. \gan{s} are typically used for obtaining a competent generator model that generates realistic images similar to the training data. Therefore, the usual training data of \gan comprises of samples belonging to only one category. The biases inherited by the discriminator through such training are not a concern, as the discriminator usually gets discarded at the end of training. In our case, we wish to retain the discriminator as a robust classifier and discard the generator instead. Thus, retaining the capabilities of the original classifier is crucial. In principle, the generator is expected to exhaust all the failure regions of the input space, including those corresponding to \gn and \gl. However, since the input space of the model can be considered infinitely large for all practical purposes, a generator model starting from a random initialisation of parameters can only attack the discriminator from \emph{one side} with respect to the training data. During training, it typically covers only a limited region that lies along the path explored by it and cannot leap to a vastly different region of input space. Our experiments also confirmed that training a \gan with only the \inj dataset results in a bias towards the signal class and a loss of performance in the other noise classes, \gn and \gl. To make the training setup more complete and retain the performance of our classifier on all three classes of data, \inj, \gn and \gl, we use the same composition of training data that was used for training the classifier as described in Section~\ref{ch:RDL:sec:utility}.

\subsection{Training for improved robustness}\label{ch:RDL:sec:robustness_impl}

\begin{figure*}
        \subfloat[Branch structure.]
        {%
            \hspace{2cm}
            \includegraphics[width=0.45\textwidth]{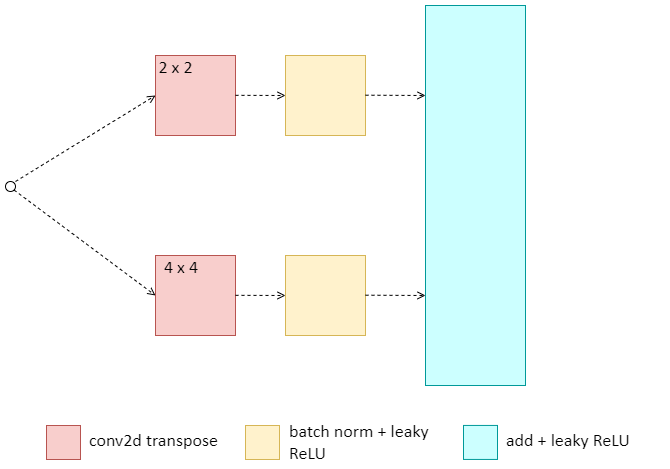}
            \vspace{0.1cm}
        } \hfill \vspace{0.1cm}
        \subfloat[Overall generator architecture.]
        {%
            \centering
            \includegraphics[width=0.8\textwidth]{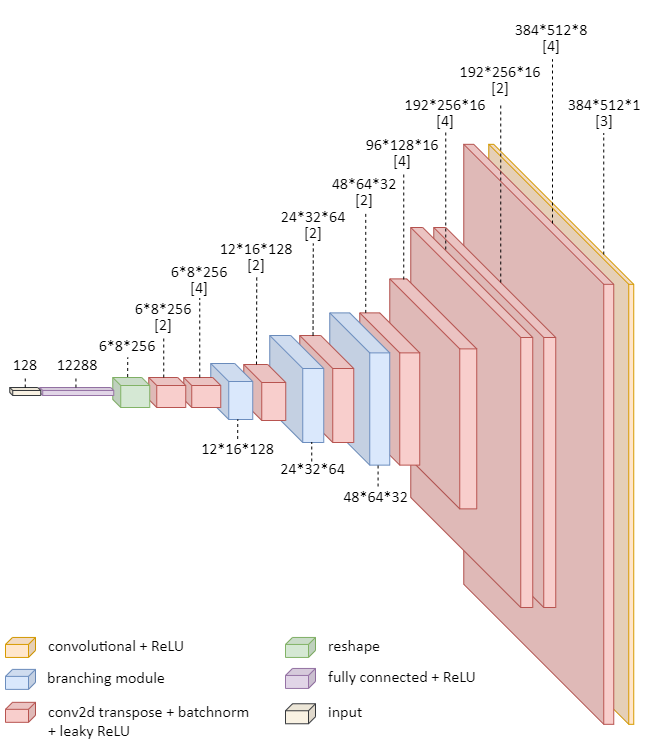}
            % \vspace{0.1cm}
        }
        \caption{Architecture of the generators used for adversarial attacks and in \gan training with their branch structure. The input to the generator is a noise vector of size $256$. The generator outputs are the fake images deceiving the classifier model.
        }
        \label{ch:RDL:fig:arch_generator}
    \end{figure*}

    As mentioned before, since covering the entire input space is practically impossible, absolute robustness is unachievable. Nevertheless, to exhaust as much space around the input space as possible, we create a setup where our fine-tuned hybrid model is simultaneously beset with $5$ generators\footnote{The number of generators considered here is limited by the computational capacity.} initialised independently. The schematic of the setup is shown in Figure~\ref{ch:RDL:fig:gan_arch}. We use three loss terms to constitute the total loss for training the discriminator. The first loss term is constructed from the fake generated samples as an average of losses obtained from each generator. The other two are \inj data loss and an average of \gn and \gl losses. The detailed architecture of the generator is shown in Figure~\ref{ch:RDL:fig:arch_generator}. We train the \gan setup for $400$ epochs (run-time $\sim 32$ hours) with a cyclical learning rate as used previously. The evolution of generator and discriminator losses through the course of training is shown in Figure~\ref{ch:RDL:fig:gan_losses}. During the initial epochs of training, the incapability of the generators leads to their relatively higher losses. However, within a few epochs, the generators start producing images with features the discriminator is possibly sensitive to. Through simultaneous training, the discriminator learns to eliminate the sensitivity to these artefacts. Eventually, the samples generated by the generators start resembling the \inj data present in the training data. 

    \begin{figure*}
        \subfloat[]
        {%
            \centering
            \includegraphics[clip=true, trim=1cm 0.7cm 1.5cm 1.8cm, width=0.49\textwidth]{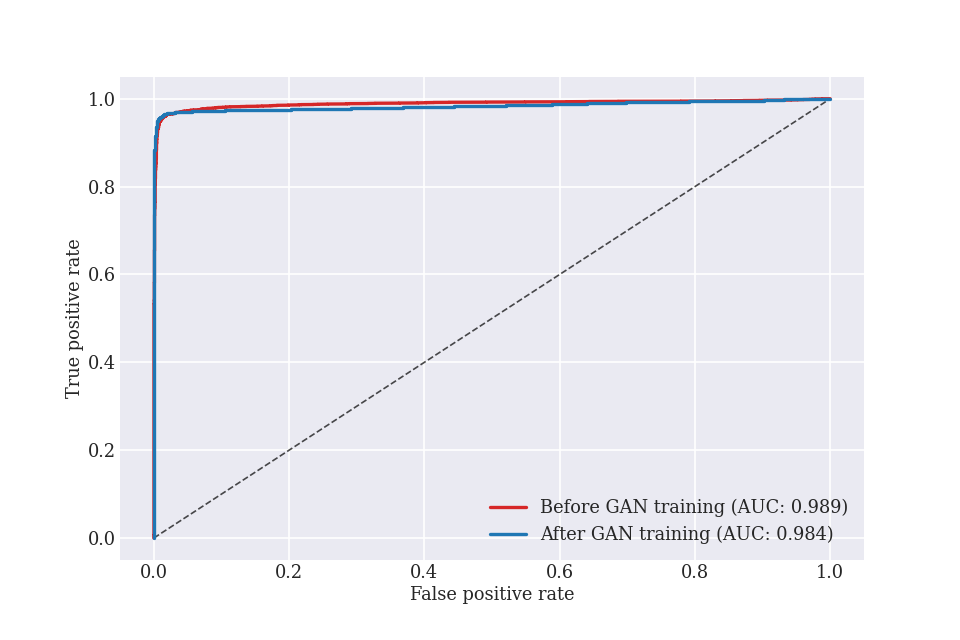}
        } \hfill
        \subfloat[]
        {%
            \centering
            \includegraphics[clip=true, trim=1cm 0.7cm 1.5cm 1.8cm, width=0.49\textwidth]{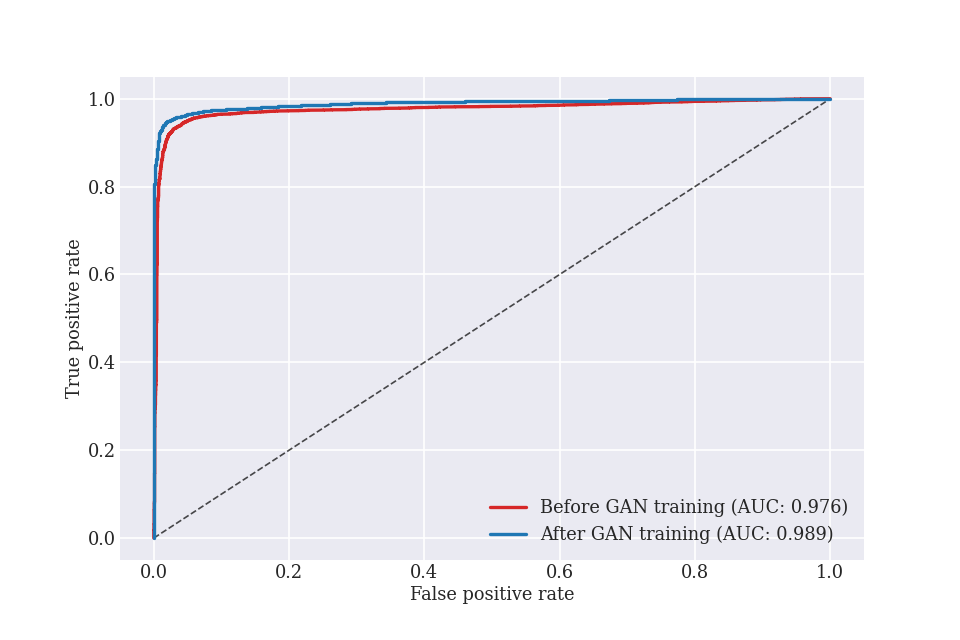}
        }
        \caption{Classifier performance on real datasets before and after the \gan training. The ROC curves on the left were obtained through inference of \injtest and \gltest as the signal and noise datasets, respectively, whereas for those on the right, the noise dataset was \trigtest. The $0.5\%$ decrease and $1.3\%$ increase in the AUC values of the two test cases are statistically insignificant, indicating that GAN training does not adversely impact the performance of the classifier on real data. Also, it is worth noting that the overall AUC values are sufficiently high, given that, except for \gl, we did not use real data during training.}
        \label{ch:RDL:fig:gan_testing}
    \end{figure*}

\begin{figure*}
        \centering
        \includegraphics[clip=true, trim=4cm 3cm 3cm 3cm, width=0.45\textwidth]{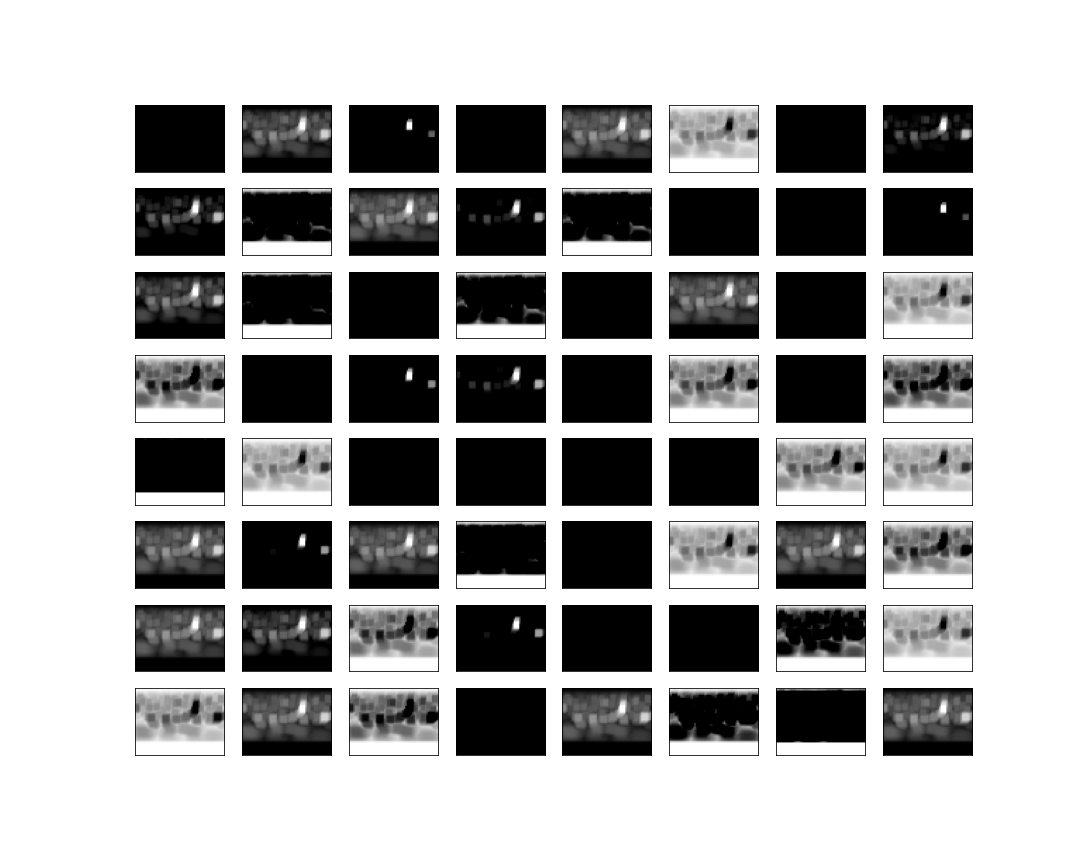}
        \includegraphics[clip=true, trim=4cm 3cm 3cm 3cm, width=0.45\textwidth]{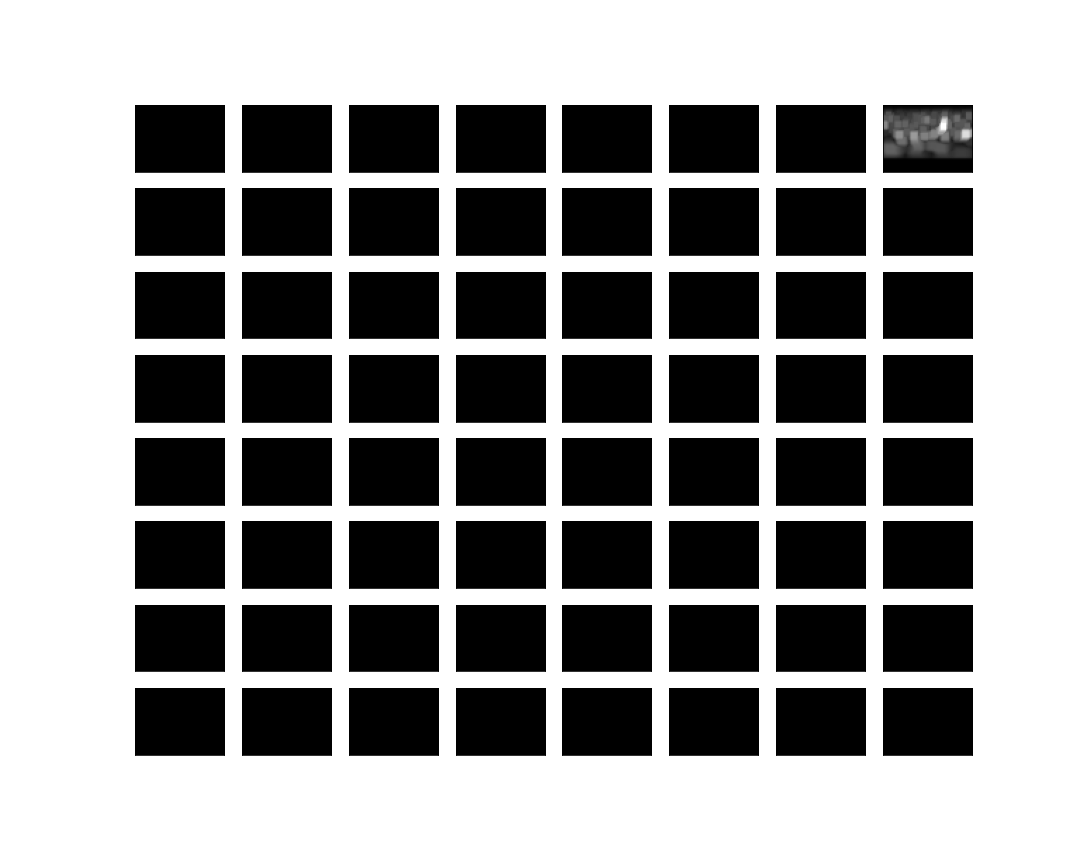}
        \includegraphics[clip=true, trim=4cm 3cm 3cm 3cm, width=0.45\textwidth]{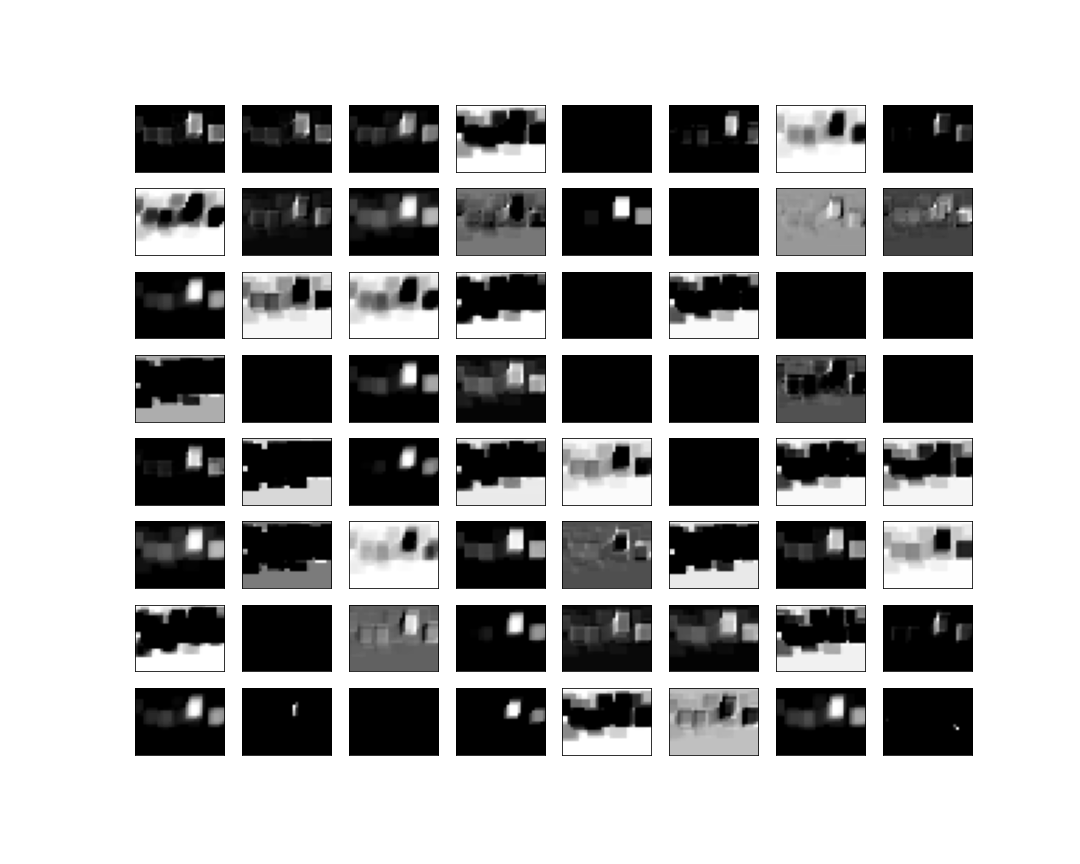}
        \includegraphics[clip=true, trim=4cm 3cm 3cm 3cm, width=0.45\textwidth]{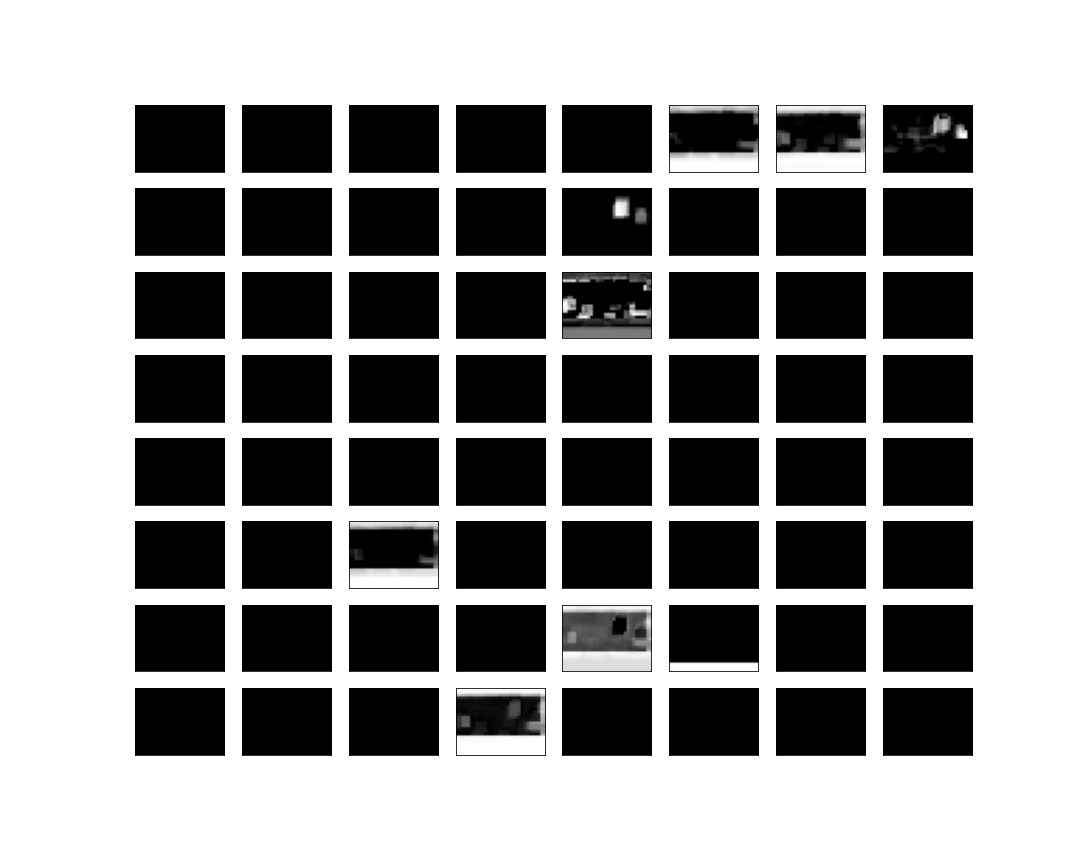}
        \includegraphics[clip=true, trim=4cm 3cm 3cm 3cm, width=0.45\textwidth]{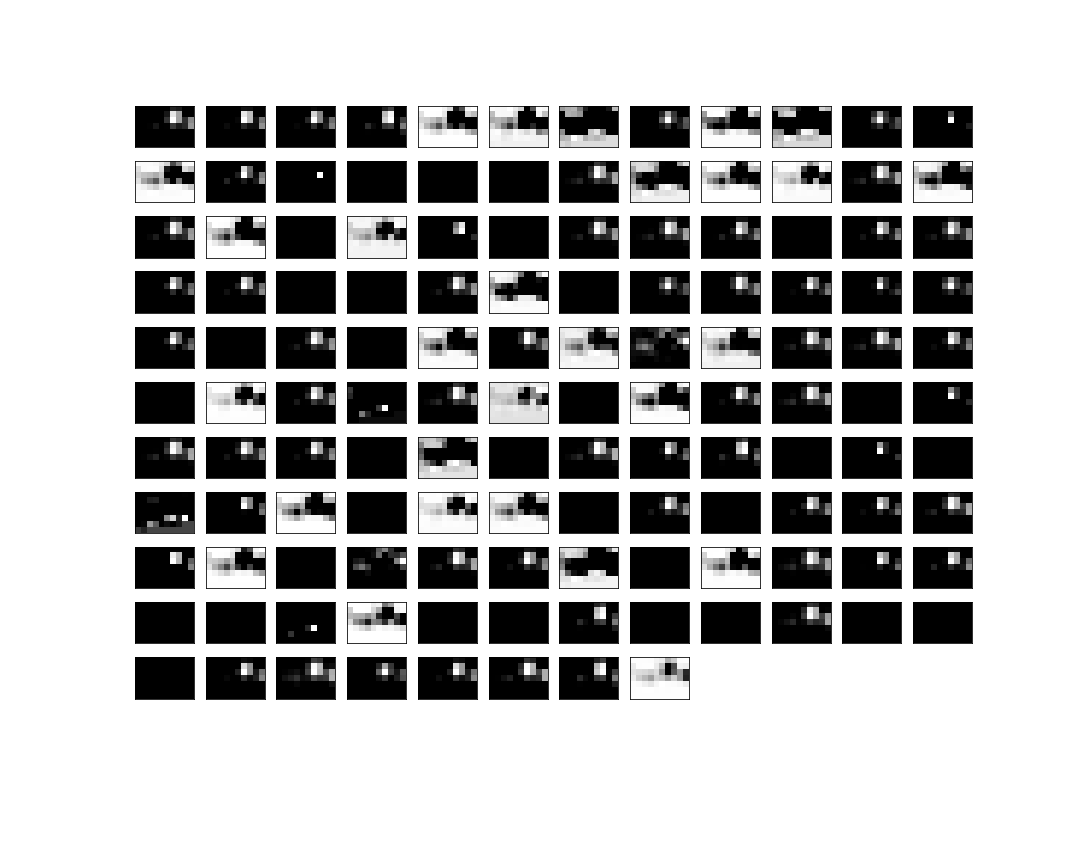}
        \includegraphics[clip=true, trim=4cm 3cm 3cm 3cm, width=0.45\textwidth]{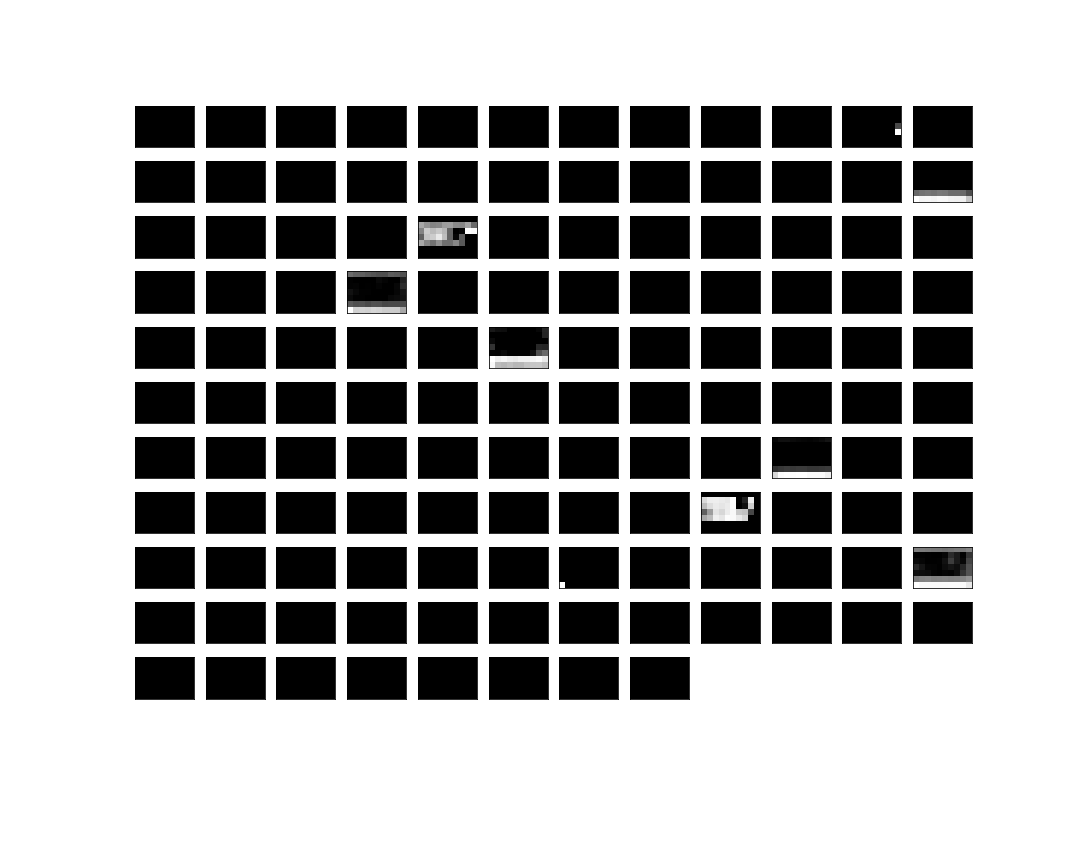}
        \caption{Comparison of image transformations of a sample \cwt map taken from the \inj database with the model-state before (left column) and after the \gan training (right column). Outputs of the model extracted at the end of the $3 \times 3$ convolutional layers from second (top) and third parallel modules (middle), and the one before flattening operation (bottom) are shown. The extracted features on the left show many similarities. On the right, these degeneracies disappear, and only a few more helpful sets of features are extracted, which can be mapped to the binary output more reliably.}
        \label{ch:RDL:fig:inj_transform_layers}
    \end{figure*}

\begin{figure*}
        \centering
        \includegraphics[clip=true, trim=4cm 3cm 3cm 3cm, width=0.45\textwidth]{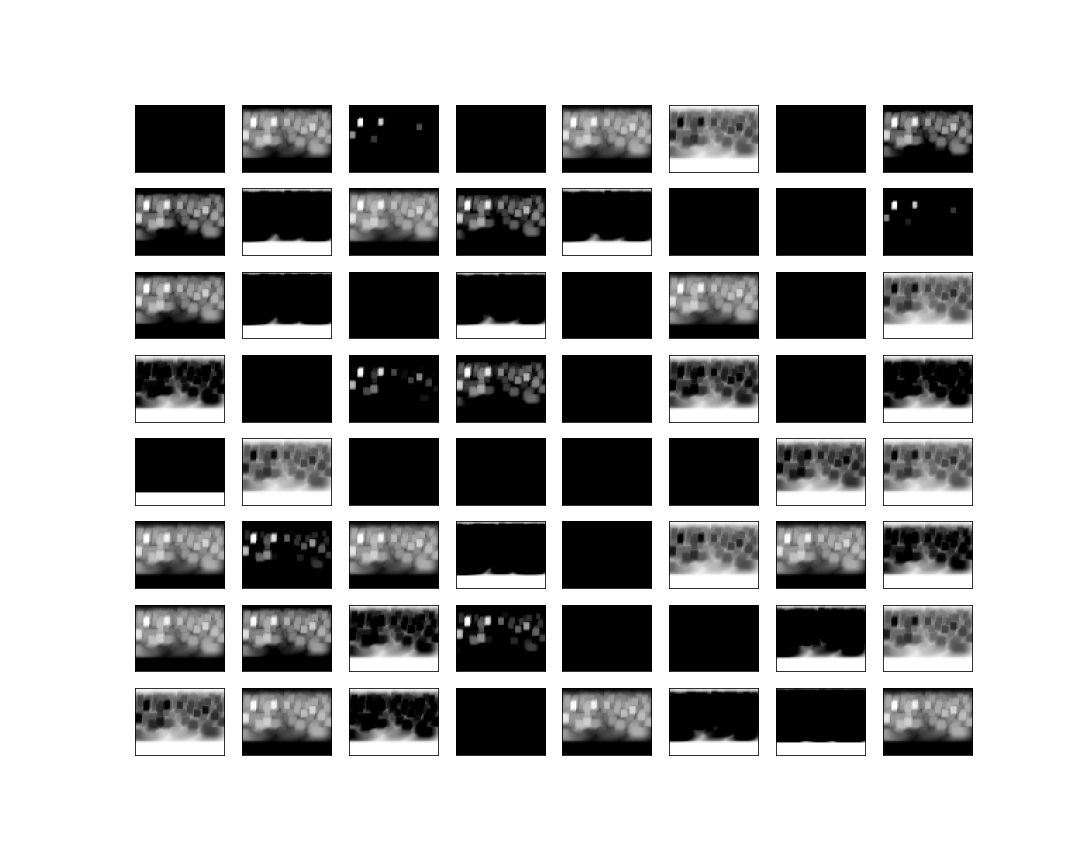}
        \includegraphics[clip=true, trim=4cm 3cm 3cm 3cm, width=0.45\textwidth]{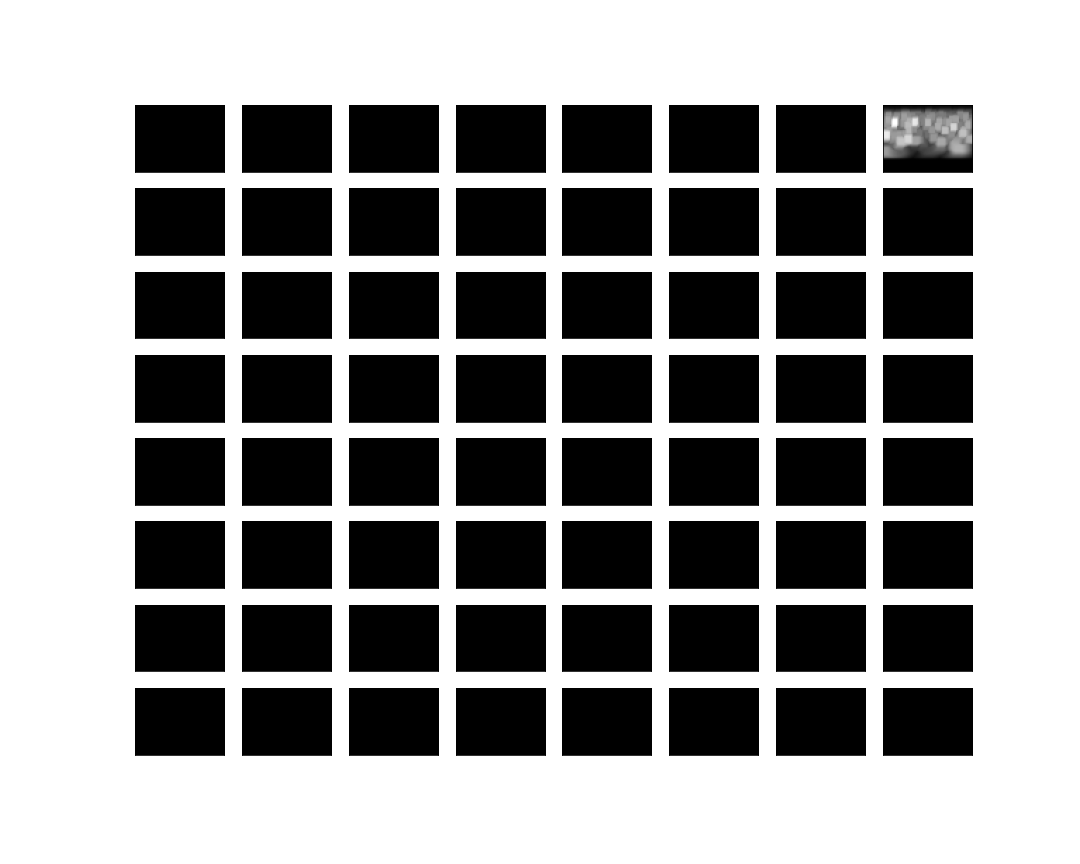}
        \includegraphics[clip=true, trim=4cm 3cm 3cm 3cm, width=0.45\textwidth]{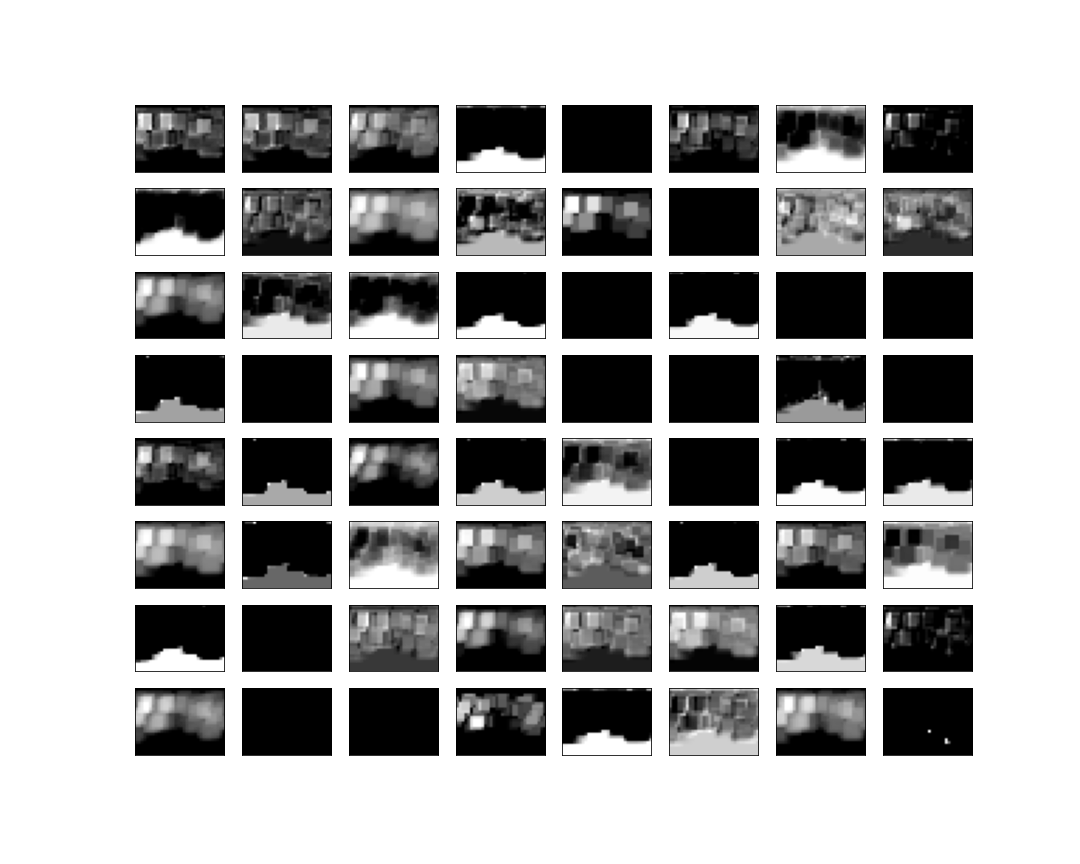}
        \includegraphics[clip=true, trim=4cm 3cm 3cm 3cm, width=0.45\textwidth]{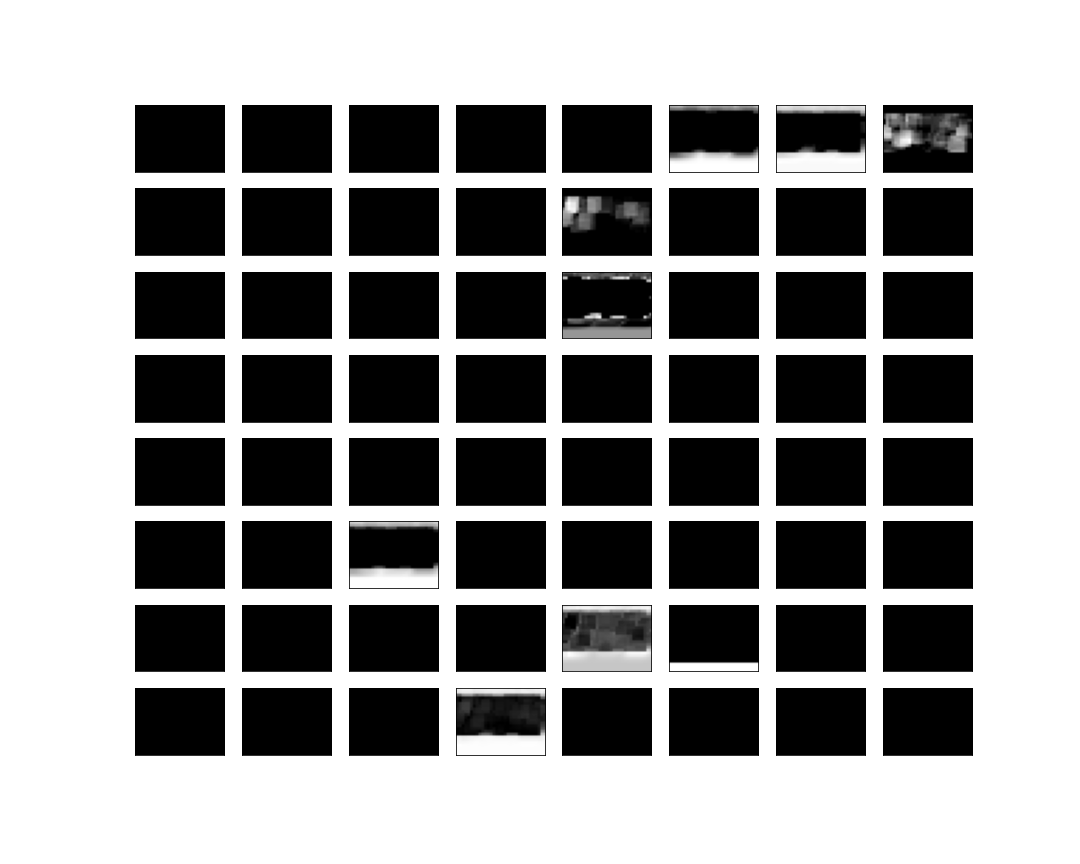}
        \includegraphics[clip=true, trim=4cm 3cm 3cm 3cm, width=0.45\textwidth]{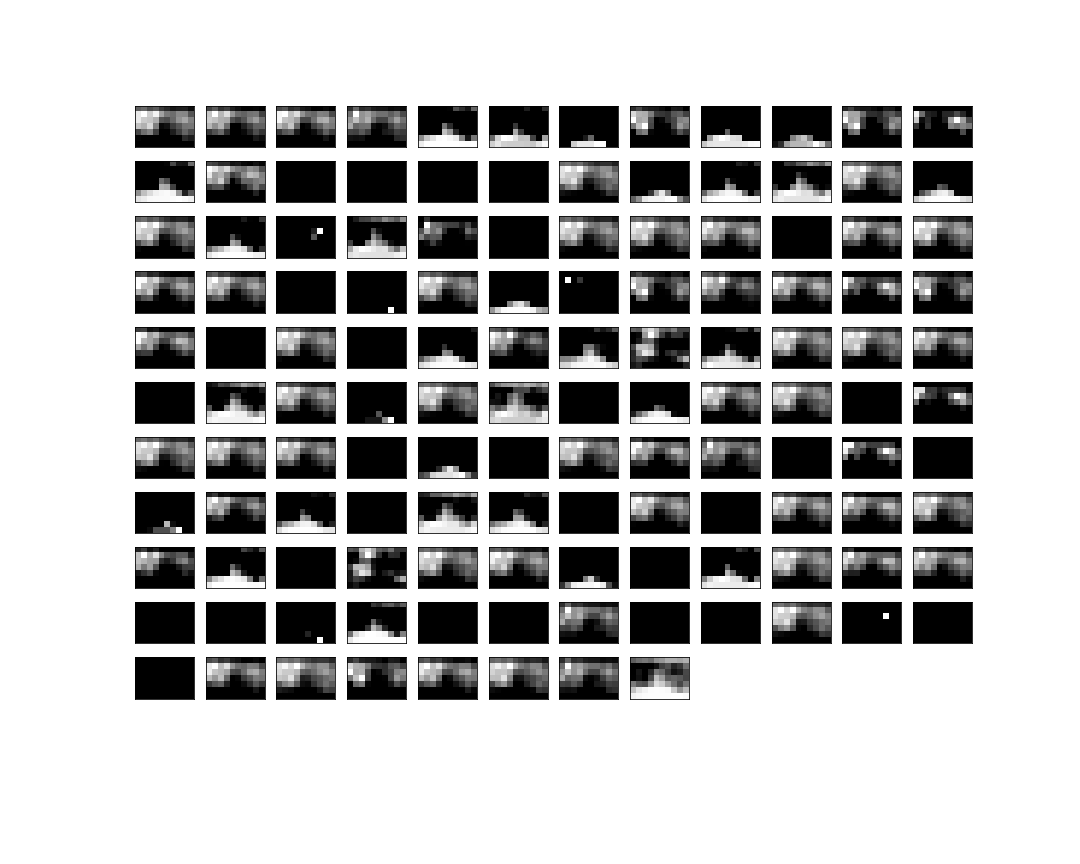}
        \includegraphics[clip=true, trim=4cm 3cm 3cm 3cm, width=0.45\textwidth]{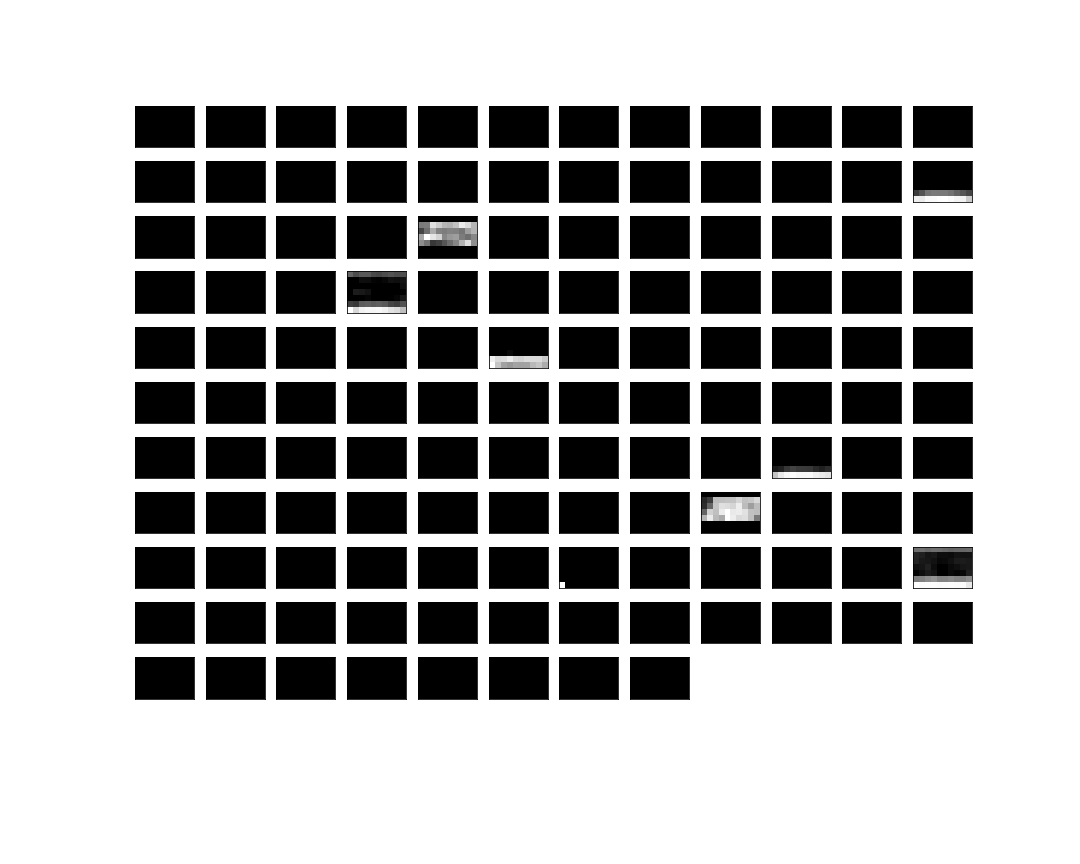}
        \caption{Comparison of image transformations similar to Figure~\ref{ch:RDL:fig:inj_transform_layers} with a sample \cwt map taken from the \gn database. A similar difference in feature extraction can be observed between the two models.}
        \label{ch:RDL:fig:noise_transform_layers}
    \end{figure*}

    Since it is impossible to achieve absolute robustness using this approach, the very need for such an additional training effort may naturally be questioned. To thoroughly examine the actual changes the \gan training brought about, we inspect the transformations of sample input data inside our classifier at various levels with and without \gan training. Specifically, we focus on the changes in the operations of $3 \times 3$ convolutional filters from the second and third parallel modules, and the one before the flattening operation towards the end. We pass a sample image from \inj and \gn data each through the classifier states before and after \gan training and obtain the transformed maps at the outputs of these convolutions as shown in Figure~\ref{ch:RDL:fig:inj_transform_layers} and Figure~\ref{ch:RDL:fig:noise_transform_layers}. For the classifier state before \gan training, the transformed maps display a high degree of similarity in terms of the features collected from the input image. Whereas, in the case of the robust classifier obtained after the \gan training, this degeneracy is largely eliminated. Fewer and more helpful features are recovered that can be translated to the binary output more reliably. This can be compared with the sparseness achieved through regularisation~\cite{Goodfellow_DL} which is a highly desirable property for a \dl model.

    To test the performance of the classifier on real data and study the possible deterioration in its capabilities after \gan training, we ran the inference on the test datasets with the states of the classifier before and after \gan training. Figure~\ref{ch:RDL:fig:gan_testing} shows the separate sets of \roc curves with \gltest and \trigtest as the noise class for classifier before and after \gan training. The \auc corresponding to the \roc curve of \cbc{s} vs glitches reduces from $0.989$ to $0.984$, while for \cbc{s} vs \pycbc triggers it improves from $0.976$ to $0.989$ after \gan training. This change is relatively insignificant and can be attributed to statistical fluctuations in performance. Therefore, we can say that the improvement in robustness is achieved with no adverse impacts in terms of the model performance on real data.

\section{Search}\label{ch:RDL:sec:search}

    Though improving the sensitivity of \cbc searches was not the target of this work, to assess the capabilities of our \gan trained model to identify events in real data, we implement a direct search on the \gw strain data from H1 and L1\footnote{Though we have not used the data from Virgo in this work, our analysis is extendable to include data from multiple detectors.}. For inference, we analyse the \gw strain data in $512$ seconds long chunks. The data were first whitened with a \psd generated from the same chunk. We analyse $1$ second-long slices of whitened strain by obtaining their \cwt maps. For analysis, we stride the data with fixed steps, generate \cwt maps and save them for inference with \gpu later. Taking a shorter stride leads to wasteful computation of image generation and \gpu inference, whereas taking longer strides leads to the possibility of missing out on signals. To systematically decide a reliable stride value, we need to identify a sensitive time window of the classifier. One can anticipate the sensitive time window as a function of chirp mass since the chirp morphology is expected to affect such a window. Once this window function is obtained, we can choose an optimal stride step such that no detectable signal is missed while the number of images generated and analysed remains as low as possible.

\subsection{Estimating the sensitive time-window}\label{ch:RDL:sec:sensitive_window}

    \begin{figure}
        \centering
        \includegraphics[width=\linewidth]{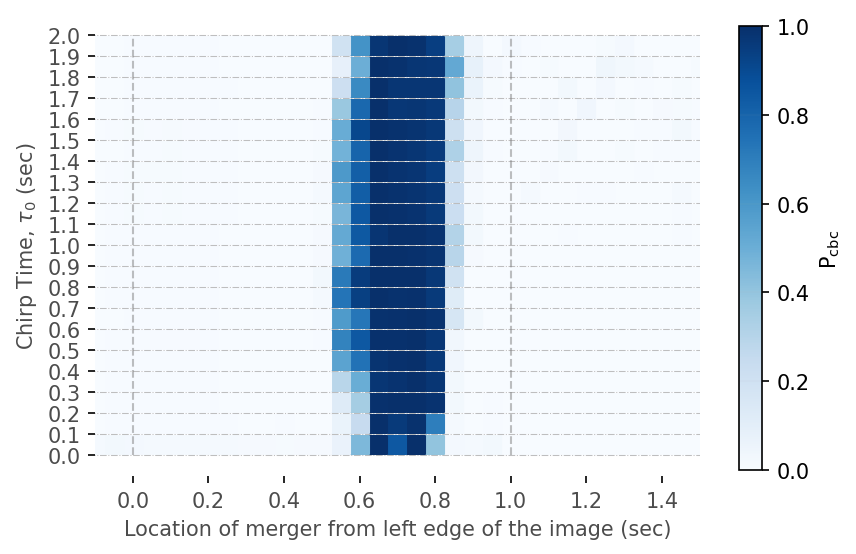}
        \caption{The sensitivity of the classifier as a function of the location of the merger in the \cwt map and the chirp-time of the \cbc signal. The injections are stridden in steps of $0.05$ second across the analysis window of $1$ second (shown by dotted lines) and the average \Pcbc values are shown with a colour bar. The sensitive window can be seen to become relatively narrower for injections with shorter duration.}
        \label{ch:RDL:fig:sensitive_window}
    \end{figure}

    To obtain the sensitive time window of the classifier, we produce a set of injections generated by uniformly sampling the chirp time $\tau_0$ between $0.01-2$ seconds. To calculate $\tau_0$, we take the starting frequency as $30$ Hz. The component masses are constrained to lie in the range $2-400\ \msun$ with mass ratio $q < 8$. The injections were generated without noise, in a setting similar to the preparation of \cleaninj dataset as described in Section~\ref{ch:RDL:sec:dbprep}. Next, the analysis window was stridden over the injection so that the merger of the signal traversed from the left to right end of the window in steps of $0.05$ second. The complex \cwt map of the clean injection was obtained at each step and added to the \cwt maps of $10$ different simulated noise realisations. An average of the predictions on this stack of \cwt maps gives a more reliable output score which was recorded for the corresponding injection at the respective step. The injections were generated with optimal \snr values such that the corresponding $\gamma$ values came out to be $50$ based on the respective component masses. The \texttt{aLIGOZeroDetHighPower} \psd was used for generating simulated noise and calculating the optimal \snr. The resulting response function obtained from the classifier is shown in Figure~\ref{ch:RDL:fig:sensitive_window}. It can be observed that the sensitive window is $\sim 0.2$ second for signals with higher $\tau_0$ values, whereas it reduces to $\sim 0.1$ second for the shortest signals. 
    Based on this plot, we heuristically chose the stride value to be $0.1$ second.

\subsection{Search on real LIGO data}\label{ch:RDL:sec:search_real_data}

    \begin{table*}
        \setlength{\tabcolsep}{12pt}
        \centering
        \begin{tabular}{l l}
            \hline
            Parameter & Value \\
            \hline
            Chirp-distance & Uniform in ($5$, $400$) Mpc \\
            Component masses & log($m_{1,2}$) uniform in $(2.5, 150)\ \msun$\\
            Mass ratio & $q$ allowed in $(1, 8)$\\
            Component spins & Aligned, $S_{i,z}$ uniform in ($-0.998$, $0.998$) \\
            Inclination & Uniform in cos$^{-1}\iota$ \\
            Sky location & Uniform over sky \\
            Approximant & \texttt{SEOBNRv4\_opt} \\
            Injection times & Uniform in a $25$ seconds interval; Average step of $100$ seconds \\
            Low frequency cut-off & $19$ Hz \\
            \hline
        \end{tabular}
        \caption{Parameter choices for the injections in the real LIGO data used for sensitivity estimation of the direct \dl search.}
        \label{ch:RDL:tab:inj_testing}
    \end{table*}

    \begin{figure}
        \centering
        \includegraphics[width=\linewidth]{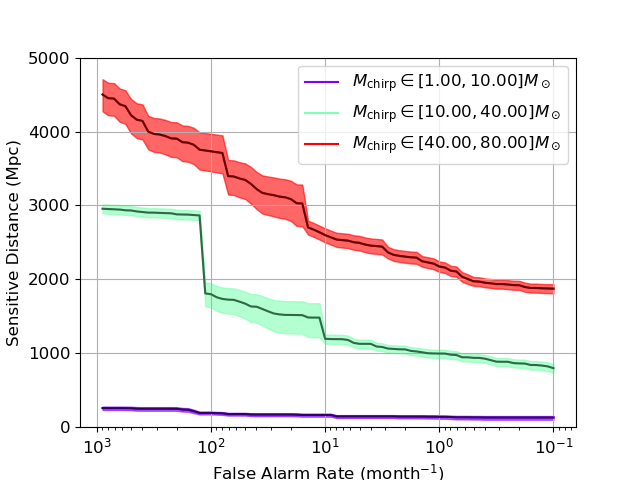}
        \caption{The sensitive distance of the \dl search as a function of FAR for three chirp-mass bins.}
        \label{ch:RDL:fig:search_sensitivity}
    \end{figure}

    \begin{figure}
        \centering
        \includegraphics[width=\linewidth]{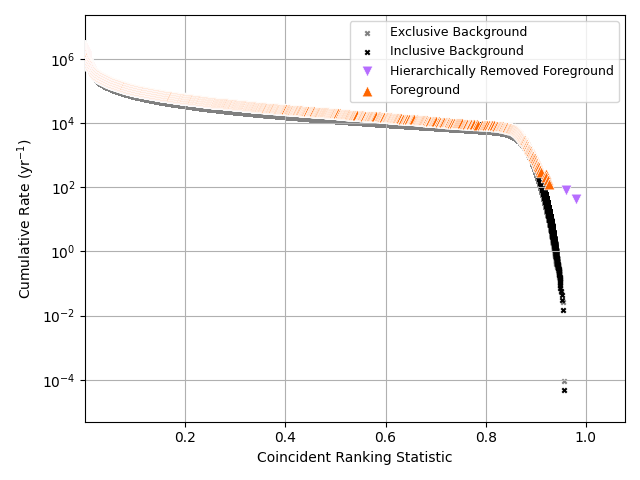}
        \caption{Cumulative rate of foreground and background events (inclusive and exclusive of foreground triggers) plotted against the coincident ranking statistic. The direct \dl search was performed on $8.8$ days of real data from LIGO Hanford and Livingston. Two \cbc events, GW190519\_153544 and GW190521\_074359, were recovered with \far $< 1$ per $20350.4$ years.
        }
        \label{ch:RDL:fig:snrifar}
    \end{figure}

    To test the independent performance of our classifier, we perform a \cbc search on $8.8$ days long chunk of real data from LIGO Hanford and Livingston detectors between May $12-21$, $2019$ UTC. This data includes six gravitational wave candidates with $\mathrm{P}_{astro} \ge 0.5$ as listed in GWTC-2.1~\cite{gwtc2.1}. For estimating the sensitivity of the search, we also create an injection set whose parameters are described in Table~\ref{ch:RDL:tab:inj_testing}. The image data for \cwt maps were generated as described in the previous section with a stride of $0.1$ second, which translates to $\sim 5$ million images in H1 and $\sim 5.9$ million images in L1. The computational cost was observed to be $\sim 1$ CPU core second per image, out of which, a significant fraction came from the operation of saving the image file on the disk. We plan to optimise this cost in future work. We ran a multi-\gpu inference on these images to obtain the model predictions as time-series data (run-time $\sim 16$ hours on four NVIDIA Tesla P100 \glspl{gpu}). We call these predictions \Pcbc scores which correspond to how the classifier ranks the triggers for them to belong to the \cbc class. However, note that this is not a realistic statistical probability of the trigger belonging to the \cbc class. 
    Triggers were generated using peak finding and clustering over a $1$ second window. Also, a lower cut-off of $10^{-5}$ was used for trigger collection. By dividing all the triggers by the maximum trigger value of the entire run, we normalised them to the range $[0-1]$. We used the \pycbc workflow for trigger collection and post-processing of these triggers for coincident analysis. The coincident foreground events were generated by collecting triggers that lie in a time window of $0.4$ second to conservatively collect triggers captured at an offset of $\sim 0.2$ second from the actual merger time in each detector. The \Pcbc data from H1 were time-shifted in $0.9$ second steps with respect to those from L1, and the same coincidence collection procedure was repeated to obtain the background events. We build the coincident ranking statistic as a simple multiplication of the \Pcbc values of the coincident triggers, which essentially expects the triggers to belong to the \cbc class in both detectors at the same time. We do not consider single detector events in this work. We perform injection recoveries by looking for coincident foreground events lying within $2$ seconds on either side of the injections. From the significance of these recovered injections, the sensitive distance for the search can be found~\cite{Usman2016}. The sensitive distance as a function of \far is plotted in Figure~\ref{ch:RDL:fig:search_sensitivity} for different chirp-mass bins. The sensitive distance at the \far of $1$ per month is found to be $140$ Mpc, $996$ Mpc and $2173$ Mpc for chirp mass bins $[1, 10]$, $[10, 40]$ and $[40, 80]$, respectively.

    The foreground and background events obtained from the search are shown in Figure~\ref{ch:RDL:fig:snrifar}. Our search recovered two significant \cbc events, GW190519\_153544 and GW190521\_074359, with \far $< 1$ per $20350.4$ years. There are four additional events in the analysed chunk that were found by other search pipelines in GWTC-2.1, viz., GW190513\_205428, GW190514\_065416, GW190517\_055101 and GW190521. Initial investigations suggest that there could be several reasons our search did not find them, viz., one of the detectors having a weaker signal or the presence of a glitch artefact in addition to signal (GW190514\_065416 and GW190513\_205428), signal lying outside the parameter space considered for training the model (GW190521) and lack of almost any tail feature in the chirp morphology (could be due to high spin modulations in the signal) leading to resemblance with blip glitches (GW190517\_055101). However, it should be noted that due to the inherent opaqueness of the \dl models, it is very difficult to provide a definitive and complete explanation of the case-to-case performance.
    It is also worth mentioning that the search results could be further improved by optimising the hyper-parameters of the search. However, we did not pursue this aspect as improving the search sensitivity is not the goal of this work.

\section{Conclusions}\label{ch:RDL:sec:concl}

    \dl algorithms hold much promise due to their ability to learn and model vast amounts of complex data - a task that is formidably difficult to perform using classical methods. Their human-like capability to generalise the learned features, along with the deployability on accelerated computing hardware like \glspl{gpu}, make them better placed for solving many data-intensive problems across domains. However, these algorithms lack transparency over the learned features and a prescription of where they could possibly fail. The fragility of large \dl models has been exposed using various methods, and improving their robustness is an area of active research~\cite{goodfellow2014explaining,GANdef,RobGAN}.

    In the field of \gw astronomy, identifying true \gw signals in the data against the spurious noise transients of terrestrial origin is one of the leading applications of \dl that has invoked a lot of interest. However, the lack of transparency and trustworthiness of these models has been a significant hurdle in deploying them for production-level analysis and attestation of their results. In this work, we take the first steps towards having better control over the performance of the \dl models and improving their reliability to perform \cbc searches in the \gw data. The key developments are summarised below - 

    \begin{itemize}

    \item \textbf{New metric for enhanced purity of data}: We started by addressing ways to maintain the purity of training data, especially for the \cbc class across the parameter space. Due to varying chirp-time of the signals based on their component masses, the \cbc injection data generated with a constant lower threshold on \snr leads to variations in the \emph{visibility} of their chirp-features. We built a metric that accounts for variations in the peak intrinsic amplitude of the \cbc signals and enables the generation of large amounts of data with uniform visibility of chirp features across the parameter space. 

    \item \textbf{\dl model development focused on signal space}: To better control which features of the input image the model learns to focus on, we trained a \vae to reproduce only the chirp features from the input data and ignore everything else. Such a \vae encodes the signal-specific information from the input into a reduced, smooth latent space representation. Using densely connected layers, we mapped this information-rich latent coding to the binary output that corresponds to the signal or noise class. The signal class consisted of \cbc signals sampled from a focused \bbh parameter space injected into the simulated LIGO noise. Whereas, the noise class consisted of simulated Gaussian noise and glitches obtained from real LIGO data. We fine-tuned the model further to achieve high training and validation accuracies. 

    \item \textbf{Investigation of fragility of \dl model and proposal for robustness}: Going beyond just careful model building, we tested this model for robustness by subjecting it to adversarial attacks. These attacks revealed several simple failure modes of the model which did not have any features similar to \cbc signals. These results also underlined the crucial fact that despite various advantages, \dl approaches can be extremely fragile. As the first step towards bringing robustness to our \dl model, we employed a novel \gan setup to retrain the model. We used our classifier as a discriminator and trained it alongside $5$ adversarial generators. We tested the model's performance before and after \gan training on separate test datasets. These datasets were composed of injections in real data, glitches and random triggers collected from matched-filter searches. We found that the \gan training had no significant effect on the model's performance on real data, while it brought some fundamental improvements in terms of sparseness which could contribute to enhanced robustness. We obtained the transformations of sample inputs at different layers of the classifier and found that a large amount of degeneracy in the extracted features was removed after \gan training and much fewer channels were used to make inferences. Also, the outputs from the layers were better localised, which could help in enabling a more robust mapping to the binary output. 

    \item \textbf{Implementation for the search of real data and estimation of search sensitivity}: Lastly, we examined the stand-alone capability of the model to search for \cbc signals in $8.8$ days of real LIGO data. We chose an optimum stride for the generation and analysis of \cwt maps by systematically investigating the model's response to injections as a function of time. This investigation helped in optimising the computational cost while ensuring recovery of all underlying signals to which the classifier is sensitive. We also performed an injection study to estimate the sensitive distance of the \cbc search. At a \far of $1$ per month, the sensitive distance for chirp mass bins $[1, 10]$, $[10, 40]$ and $[40, 80]$ was $140$ Mpc, $996$ Mpc and $2173$ Mpc, respectively. Importantly, our search recovered two \cbc events, GW190519\_153544 and GW190521\_074359, with high significance. The analysed data also contained four additional events, GW190513\_205428, GW190514\_065416, GW190517\_055101 and GW190521, found by various pipelines in GWTC-2.1~\cite{gwtc2.1}. We discuss the potential causes of our classifier missing these signals in Section~\ref{ch:RDL:sec:search_real_data}. As search sensitivity was not the focus of this work, we did not pursue possible improvements in the results by optimising the hyper-parameters of the search. Also, the overall search sensitivity can be considerably improved if our classifier is used in conjunction with an existing offline search using \mlstat~\cite{Jadhav_2021}. In future, we intend to analyse the real LIGO/Virgo data with this classifier in the framework of \mlstat and study the improvement in the significance of events from GWTC-2 and GWTC-3~\cite{Abbott2021,gwtc3}.

    \end{itemize}

    Apart from providing speedy inference, the fast-growing field of \dl applications in \gw astronomy is expected to compete with, if not complement, conventional algorithms in terms of sensitivity and parameter space coverage in the near future. Through this work, we have addressed a crucial aspect of this field: the reliability of \dl approaches. While the novel proposal to increase the robustness presented here is still in its nascent stages and under further development, we believe these techniques will pave the way for the successful integration of \dl methods in production-level analyses. We hope this work goes beyond the field of \gw astronomy and proves to be a valuable stepping-stone for other areas of science and industrial applications.\\

\begin{acknowledgments}
Authors express thanks to the members of LIGO-Virgo-KAGRA Collaboration for their valuable comments and suggestions. SJ acknowledges support of Council for Scientific and Industrial Research (CSIR), India and the Australian Research Council Centre of Excellence for Gravitational Wave Discovery (OzGrav), project number CE170100004. SM acknowledges support from the Department of Science and Technology (DST), India, provided under the Swarna Jayanti Fellowships scheme. This material is based upon work supported by NSF's LIGO Laboratory which is a major facility fully funded by the National Science Foundation. We acknowledge the use of GWA and the LDG clusters at Inter-University Centre for Astronomy and Astrophysics (Sarathi) for the computational work. We also acknowledge National Supercomputing Mission (NSM) for providing computing resources of ‘PARAM Siddhi-AI’, under National PARAM Supercomputing Facility (NPSF), C-DAC, Pune and supported by the Ministry of Electronics and Information Technology (MeitY) and Department of Science and Technology (DST), Government of India. This document has been assigned \mbox{IUCAA} preprint number IUCAA-03/2023 and LIGO document number LIGO-P2200351.
\end{acknowledgments}

\bibliography{main}

\end{document}